\input harvmac
\edef\cite{\the\catcode`@}%
\catcode`@ = 11
\let\@oldatcatcode = \cite
\chardef\@letter = 11
\chardef\@other = 12
%
%
%
%
\def\@innerdef#1#2{\edef#1{\expandafter\noexpand\csname #2\endcsname}}%
%
%
\@innerdef\@innernewcount{newcount}%
\@innerdef\@innernewdimen{newdimen}%
\@innerdef\@innernewif{newif}%
\@innerdef\@innernewwrite{newwrite}%
%
%
%
\def\@gobble#1{}%
%
%
%
\ifx\inputlineno\@undefined
   \let\@linenumber = \empty 
\else
   \def\@linenumber{\the\inputlineno:\space}%
\fi
%
%
%
\def\@futurenonspacelet#1{\def\cs{#1}%
   \afterassignment\@stepone\let\@nexttoken=
}%
\begingroup 
\def\\{\global\let\@stoken= }%
\\ 
\endgroup
\def\@stepone{\expandafter\futurelet\cs\@steptwo}%
\def\@steptwo{\expandafter\ifx\cs\@stoken\let\@@next=\@stepthree
   \else\let\@@next=\@nexttoken\fi \@@next}%
\def\@stepthree{\afterassignment\@stepone\let\@@next= }%
%
%
%
\def\@getoptionalarg#1{%
   \let\@optionaltemp = #1%
   \let\@optionalnext = \relax
   \@futurenonspacelet\@optionalnext\@bracketcheck
}%
%
%
\def\@bracketcheck{%
   \ifx [\@optionalnext
      \expandafter\@@getoptionalarg
   \else
      \let\@optionalarg = \empty
      \expandafter\@optionaltemp
   \fi
}%
\def\@@getoptionalarg[#1]{%
   \def\@optionalarg{#1}%
   \@optionaltemp
}%
%
%
%
\def\@nnil{\@nil}%
\def\@fornoop#1\@@#2#3{}%
\def\@for#1:=#2\do#3{%
   \edef\@fortmp{#2}%
   \ifx\@fortmp\empty \else
      \expandafter\@forloop#2,\@nil,\@nil\@@#1{#3}%
   \fi
}%
\def\@forloop#1,#2,#3\@@#4#5{\def#4{#1}\ifx #4\@nnil \else
       #5\def#4{#2}\ifx #4\@nnil \else#5\@iforloop #3\@@#4{#5}\fi\fi
}%
\def\@iforloop#1,#2\@@#3#4{\def#3{#1}\ifx #3\@nnil
       \let\@nextwhile=\@fornoop \else
      #4\relax\let\@nextwhile=\@iforloop\fi\@nextwhile#2\@@#3{#4}%
}%
%
%
%
\@innernewif\if@fileexists
\def\@testfileexistence{\@getoptionalarg\@finishtestfileexistence}%
\def\@finishtestfileexistence#1{%
   \begingroup
      \def\extension{#1}%
      \immediate\openin0 =
         \ifx\@optionalarg\empty\jobname\else\@optionalarg\fi
         \ifx\extension\empty \else .#1\fi
         \space
      \ifeof 0
         \global\@fileexistsfalse
      \else
         \global\@fileexiststrue
      \fi
      \immediate\closein0
   \endgroup
}%
%
%
%
%
\def\bibliographystyle#1{%
}%
\let\bibstyle = \@gobble
%
%
\let\bblfilebasename = \jobname
\def\bibliography#1{%
      \nobreak
      \@readbblfile
}%
\let\bibdata = \@gobble
%
%
\def\nocite#1{%
   \@readauxfile
   \@writeaux{\string\citation{#1}}%
}%
\@innernewif\if@notfirstcitation
%
%
\def\cite{\@getoptionalarg\@cite}%
%
%
\def\@cite#1{%
   \let\@citenotetext = \@optionalarg
   \printcitestart
   \nocite{#1}%
   \@notfirstcitationfalse
   \@for \@citation :=#1\do
   {%
      \expandafter\@onecitation\@citation\@@
   }%
   \ifx\empty\@citenotetext\else
      \printcitenote{\@citenotetext}%
   \fi
   \printcitefinish
}%
\def\@onecitation#1\@@{%
   \if@notfirstcitation
      \printbetweencitations
   \fi
   \expandafter \ifx \csname\@citelabel{#1}\endcsname \relax
      \if@citewarning
         \message{\@linenumber Undefined citation `#1'.}%
      \fi
      \expandafter\gdef\csname\@citelabel{#1}\endcsname{%
         {\tt
            \escapechar = -1
            \nobreak\hskip0pt
            \expandafter\string\csname#1\endcsname
            \nobreak\hskip0pt
         }%
      }%
   \fi
   \csname\@citelabel{#1}\endcsname
   \@notfirstcitationtrue
}%
%
%
\def\@citelabel#1{b@#1}%
%
%
\def\@citedef#1#2{\expandafter\gdef\csname\@citelabel{#1}\endcsname{#2}}%
%
%
%
\def\@readbblfile{%
   \ifx\@itemnum\@undefined
      \@innernewcount\@itemnum
   \fi
   \begingroup
      \ifx\begin\@undefined
         \def\begin##1##2{%
            \setbox0 = \hbox{\biblabelcontents{##2}}%
            \biblabelwidth = \wd0
         }%
         \let\end = \@gobble 
      \fi
      %
      %
      \@itemnum = 0
      \def\bibitem{\@getoptionalarg\@bibitem}%
      \def\@bibitem{%
         \ifx\@optionalarg\empty
            \expandafter\@numberedbibitem
         \else
            \expandafter\@alphabibitem
         \fi
      }%
      \def\@alphabibitem##1{%
         \expandafter \xdef\csname\@citelabel{##1}\endcsname {\@optionalarg}%
         \ifx\biblabelprecontents\@undefined
            \let\biblabelprecontents = \relax
         \fi
         \ifx\biblabelpostcontents\@undefined
            \let\biblabelpostcontents = \hss
         \fi
         \@finishbibitem{##1}%
      }%
      \def\@numberedbibitem##1{%
         \advance\@itemnum by 1
         \expandafter \xdef\csname\@citelabel{##1}\endcsname{\number\@itemnum}%
         \ifx\biblabelprecontents\@undefined
            \let\biblabelprecontents = \hss
         \fi
         \ifx\biblabelpostcontents\@undefined
            \let\biblabelpostcontents = \relax
         \fi
         \@finishbibitem{##1}%
      }%
      \def\@finishbibitem##1{%
         \biblabelprint{\csname\@citelabel{##1}\endcsname}%
         \@writeaux{\string\@citedef{##1}{\csname\@citelabel{##1}\endcsname}}%
         \ignorespaces
      }%
      %
      %
      \let\em = \bblem
      \let\newblock = \bblnewblock
      \let\sc = \bblsc
      \frenchspacing
      \clubpenalty = 4000 \widowpenalty = 4000
      \tolerance = 10000 \hfuzz = .5pt
      \everypar = {\hangindent = \biblabelwidth
                      \advance\hangindent by \biblabelextraspace}%
      \bblrm
      \parskip = 1.5ex plus .5ex minus .5ex
      \biblabelextraspace = .5em
      \bblhook
      \input \bblfilebasename.bbl
   \endgroup
}%
%
%
\@innernewdimen\biblabelwidth
\@innernewdimen\biblabelextraspace
%
%
%
\def\biblabelprint#1{%
   \noindent
   \hbox to \biblabelwidth{%
      \biblabelprecontents
      \biblabelcontents{#1}%
      \biblabelpostcontents
   }%
   \kern\biblabelextraspace
}%
%
%
%
\def\biblabelcontents#1{{\bblrm [#1]}}%
%
%
\def\bblrm{\rm}%
%
%
\def\bblem{\it}%
%
%
\def\bblsc{\ifx\@scfont\@undefined
              \font\@scfont = cmcsc10
           \fi
           \@scfont
}%
%
%
\def\bblnewblock{\hskip .11em plus .33em minus .07em }%
%
%
\let\bblhook = \empty
%
%
%
\def\printcitestart{[}
\def\printcitefinish{]}
\def\printbetweencitations{, }
\def\printcitenote#1{, #1}
%
%
%
\let\citation = \@gobble
%
%
%
\@innernewcount\@numparams
%
%
\def\newcommand#1{%
   \def\@commandname{#1}%
   \@getoptionalarg\@continuenewcommand
}%
%
%
\def\@continuenewcommand{%
   \@numparams = \ifx\@optionalarg\empty 0\else\@optionalarg \fi \relax
   \@newcommand
}%
%
%
\def\@newcommand#1{%
   \def\@startdef{\expandafter\edef\@commandname}%
   \ifnum\@numparams=0
      \let\@paramdef = \empty
   \else
      \ifnum\@numparams>9
         \errmessage{\the\@numparams\space is too many parameters}%
      \else
         \ifnum\@numparams<0
            \errmessage{\the\@numparams\space is too few parameters}%
         \else
            \edef\@paramdef{%
               \ifcase\@numparams
                  \empty  No arguments.
               \or ####1%
               \or ####1####2%
               \or ####1####2####3%
               \or ####1####2####3####4%
               \or ####1####2####3####4####5%
               \or ####1####2####3####4####5####6%
               \or ####1####2####3####4####5####6####7%
               \or ####1####2####3####4####5####6####7####8%
               \or ####1####2####3####4####5####6####7####8####9%
               \fi
            }%
         \fi
      \fi
   \fi
   \expandafter\@startdef\@paramdef{#1}%
}%
%
%
%
%
\def\@readauxfile{%
   \if@auxfiledone \else 
      \global\@auxfiledonetrue
      \@testfileexistence{aux}%
      \if@fileexists
         \begingroup
            \endlinechar = -1
            \catcode`@ = 11
            \input \jobname.aux
         \endgroup
      \else
         \message{\@undefinedmessage}%
         \global\@citewarningfalse
      \fi
      \immediate\openout\@auxfile = \jobname.aux
   \fi
}%
%
%
\newif\if@auxfiledone
\ifx\noauxfile\@undefined \else \@auxfiledonetrue\fi
%
%
%
%
\@innernewwrite\@auxfile
\def\@writeaux#1{\ifx\noauxfile\@undefined \write\@auxfile{#1}\fi}%
%
%
%
\ifx\@undefinedmessage\@undefined
   \def\@undefinedmessage{No .aux file; I won't give you warnings about
                          undefined citations.}%
\fi
%
%
\@innernewif\if@citewarning
\ifx\noauxfile\@undefined \@citewarningtrue\fi
%
%
%
\catcode`@ = \@oldatcatcode

\def\href#1#2{#2}  
\input labeldefs.tmp
\writedefs
\sequentialequations


\def\TL{\hfil$\displaystyle{##}$}
\def\TR{$\displaystyle{{}##}$\hfil}
\def\TC{\hfil$\displaystyle{##}$\hfil}
\def\TT{\hbox{##}}



\def\comment#1{}
\def\fixit#1{}

\def\tf#1#2{{\textstyle{#1 \over #2}}}


\def\vol{\mathop{\rm vol}\nolimits}
\def\diag{\mathop{\rm diag}\nolimits}
\def\tr{\mathop{\rm tr}\nolimits}

\def\lsim{\mathrel{\mathstrut\smash{\ooalign{\raise2.5pt\hbox{$<$}\cr\lower2.5pt\hbox{$\sim$}}}}}
\def\gsim{\mathrel{\mathstrut\smash{\ooalign{\raise2.5pt\hbox{$>$}\cr\lower2.5pt\hbox{$\sim$}}}}}


\def\slashed#1{\ooalign{\hfil\hfil/\hfil\cr $#1$}}

\def\sqr#1#2{{\vcenter{\vbox{\hrule height.#2pt
         \hbox{\vrule width.#2pt height#1pt \kern#1pt
            \vrule width.#2pt}
         \hrule height.#2pt}}}}
\def\square{\mathop{\mathchoice\sqr56\sqr56\sqr{3.75}4\sqr34}\nolimits}





\def\hc{{\rm h.c.}}
\def\+{^\dagger}
\def\omicron{o}
\def\O{{\cal O}}


\def\overleftrightarrow#1{\vbox{\ialign{##\crcr
     $\leftrightarrow$\crcr\noalign{\kern-0pt\nointerlineskip}
     $\hfil\displaystyle{#1}\hfil$\crcr}}}
\def\em{\it}   




\Title{
 \vbox{\baselineskip10pt
  \hbox{PUPT-1699}
  \hbox{hep-th/9706100}
 }
}
{
 \vbox{
  \centerline{Absorption of photons and fermions by}
  \vskip 0.1 truein
  \centerline{black holes in four dimensions}
 }
}
\vskip -25 true pt

\centerline{
 Steven S.~Gubser,\footnote{$^1$}{e-mail:  ssgubser@viper.princeton.edu}}
\centerline{\it Joseph Henry Laboratories, 
Princeton University, Princeton, NJ  08544}

\centerline {\bf Abstract}
\smallskip
\baselineskip12pt
\noindent

The absorption of photons and fermions into four-dimensional black
holes is described by equations which in certain cases can be analyzed
using dyadic index techniques.  The resulting absorption
cross-sections for near-extremal black holes have a form at low
energies suggestive of the effective string model.  A coupling to the
effective string is proposed for spin-$0$ and spin-$1/2$ fields of
pure $N=4$ supergravity which respects the unbroken supersymmetry of
extreme black holes and correctly predicts dilaton, axion, and fermion
cross-sections up to an overall normalization.

\Date{June 1997}

\noblackbox
\baselineskip 14pt plus 1pt minus 1pt


\newsec{Introduction}
\seclab\Introduction

Microscopic models of near-extremal black holes in terms of effective
strings have recently been employed with great success to explain the
Bekenstein-Hawking entropy \cite{sv,cm,HS,hms,ms}.  These models have
also proven their value by correctly predicting certain Hawking
emission rates and absorption cross-sections at low energies
\cite{dmw,dmOne,dmTwo,gkOne,gkTwo,cgkt,dkt,htr,kvk,km,krt,mast,ja,clOne,clTwo}.

The many successful predictions have all been for scalar particles.
Most in fact have been for minimally coupled scalars.  Usually it is
said that minimally coupled scalars are those whose equation of motion
is $\square \phi = 0$.  This is a slight misnomer: a more precise way
to say it is that when the equations of motion are linearized in small
fluctuations around the black hole solution under consideration, one
of them is simply $\square \delta\phi = 0$.  Fortunately, it is
usually easy to see which scalars in the theory are minimally coupled:
for example, when one obtains a 5-dimensional black hole by toroidally
compactifying the D1-brane D5-brane bound state, the off-diagonal
gravitons with both indices lying within the D5-brane but
perpendicular to the D1-brane are clearly minimally coupled scalars in
the 5-dimensional theory.

As a starting point in the study of particles with nonzero spin, it is
easiest to consider minimally coupled photons or fermions.  For
photons, minimally coupled means that the relevant linearized equation
of motion is
  \eqn\MaxEq{
   \nabla_\mu \delta F^{\mu\nu} = 0 \ ,
  }
 and for chiral fermions it means
  \eqn\WeylEq{
   \sigma_\mu^{\alpha\dot\beta} \nabla^\mu \delta \psi_\alpha = 0 \ .
  }
 Minimally coupled fermions in arbitrary dimensions were studied in
\cite{dgm}.  The authors of \cite{dgm} correctly point out that the
fermions in supergravity theories do not in general obey minimally
coupled equations of motion in the presence of charged black holes.
Likewise, it is not usually the case that photons will be minimally
coupled in the presence of a supergravity black hole solution.
Indeed, it seems that the gauge fields which carry the charges of the
black hole are never minimally coupled: there is mixing between them
and the graviton.

However, for the case I shall study, the equal charge black hole
\cite{klopp} of $N=4$ supergravity \cite{adas,csfOne,cs,csfTwo}, two
of the four Weyl fermions are in fact minimally coupled, as are four
of the six gauge fields.  The same black hole provided the simplest
framework in which to study fixed scalars \cite{kr}.  Its metric is
that of an extreme Reissner-Nordstrom black hole.  In \cite{ja} it was
shown that an effective string model is capable of reproducing the
minimally coupled scalar cross-section of the Kerr-Newman metric, in
the near-extremal limit.  The effective string picture carries over
naturally to the equal charge extreme black hole in $N=4$ supergravity
and its near-extremal generalization.  The recent work \cite{clTwo}
presents evidence that the effective string can model a much broader
class of black holes which have arbitrary $U(1)$ charges and are far
from extremality.  The essential features of the effective string,
however, seem much the same in all its four-dimensional applications.
I will show that minimally coupled fermions can be incorporated
naturally into the effective string picture through a coupling to the
supercurrent.  Minimally coupled photons fit in in a somewhat
unexpected way: the coupling of the gauge field to the string seems to
occur via the field strength rather than the gauge potential.

The organization of the paper is as follows.  In section~\MinCoup, the
black hole solutions are exhibited and the minimally coupled photons
are identified.  In section~\SepEq, separable equations are derived
for these photons.  In section~\Prob, these equations are solved to
yield absorption cross-sections.  The parallel analysis of minimally
coupled fermions is postponed to section~\OtherParticles, in which
also the axion cross-section is computed.  The axion turns out to have
the same cross-section as the dilaton, not because it is a fixed
scalar in the usual sense of attractors \cite{fkOne,fkTwo}, but
because of a dynamical version of the Witten effect.  Section~\EffStr\
discusses the effective string interpretation of these cross-sections.
Although the overall normalizations of the cross-sections are not
computed, it is shown that the effective string correctly reproduces
the relative normalization of the dilaton, axion, and minimal fermion
cross-sections.  Some concluding remarks are made in
section~\Conclusion.  Appendix~A presents some results of the dyadic
index formalism needed for the rest of the paper.

\newsec{Minimally coupled photons in $N=4$ supergravity}
\seclab\MinCoup

The fields of the $SU(4)$ version of $N=4$, $d=4$ supergravity
\cite{csfTwo} are the graviton $e_\mu^a$, four Majorana gravitinos
$\psi^i_\mu$, three vector fields $A^n_\mu$, three axial vectors
$B^n_\mu$, four Majorana fermions $\chi^i$, the dilaton $\phi$, and
the axion $B$.  The doubly extreme black hole of \cite{klopp} is
electrically charged under $A_3^\mu$, magnetically charged under
$B_3^\mu$, and neutral with respect to the other four gauge fields.
These four extra gauge fields are minimally coupled photons, as we
shall see shortly.

The full bosonic lagrangian of $N=4$ supergravity in the $SU(4)$
picture is
  \eqn\FullBL{\eqalign{
    {\cal L} &= \sqrt{-g} \Big[ -\!R + 2 (\partial_\mu \phi)^2 + 
     2 e^{4 \phi} (\partial_\mu B)^2 - 
     e^{-2 \phi} {\textstyle \sum_n} (F_n^2 + G_n^2)  \cr
   &\qquad - \, 2 i B {\textstyle \sum_n} (F_n *F_n + G_n *G_n) \Big] 
  }}
 where $F_n = d A_n$ and $G_n = d B_n$.  The conventions used here are
those of \cite{klopp}.  In particular, Hodge duals are defined by
  \eqn\DualDef{
   *F_{\mu\nu} = {\sqrt{-g} \over 2} \epsilon_{\mu\nu\rho\sigma}
    F^{\rho\sigma}
  }
 where $\epsilon_{0123} = \epsilon_{tr\theta\phi} = -i$.  The
equations of motion following from \FullBL\ are
  \eqn\SEOMs{\eqalign{
   \nabla_\mu \big( e^{-2\phi} F_n^{\mu\nu} + 
     2iB *F_n^{\mu\nu} \big) &= 0  \cr 
   \nabla_\mu \big( e^{-2\phi} G_n^{\mu\nu} +
     2iB *G_n^{\mu\nu} \big) &= 0  \cr
   \square\phi - \tf{1}{2} e^{-2\phi} {\textstyle \sum_n} 
    (F_n^2 + G_n^2) - 2 e^{4\phi} (\partial_\mu B)^2 &= 0  \cr 
   \square B + 4 \partial^\mu \phi \partial_\mu B +
    \tf{i}{2} e^{-4\phi} {\textstyle \sum_n} (F_n *F_n + G_n *G_n) 
       &= 0  \cr
   R_{\mu\nu} + 2 \partial_\mu \phi \partial_\nu \phi +
    2 e^{4\phi} \partial_\mu B \partial_\nu B
    - e^{-2\phi} {\textstyle \sum_n} 
    (2 F_{n\mu\lambda} F_{n\nu}{}^\lambda - \tf{1}{2} g_{\mu\nu} F_n^2 
      \ \ \, &  \cr
     + \, 2 G_{n\mu\lambda} G_{n\nu}{}^\lambda - \tf{1}{2} g_{\mu\nu} G_n^2) 
    &= 0 \ .  
  }}
 Including the fermions introduces extra terms into these equations
involving fermion bilinears.  These terms affect neither the black
hole solution nor the linearized bosonic equations around that
solution since the fermions' background values are zero.  Duals of the
field strengths $F_n$ and $G_n$ are defined by
  \eqn\FSDuals{
   \tilde{F} = i e^{-2 \phi} * F - 2 B F \ .
  }
 $\tilde{F}_n$ and $\tilde{G}_n$ are closed forms by the equations of
motion \SEOMs.  In the $SO(4)$ version of $N=4$ supergravity, one
writes $\tilde{G}_n = d \tilde{B}_n$.

The equal charge, axion free, extreme black hole solution is
  \eqn\kpSol{\eqalign{
   ds^2 &= {1 \over (1+M/r)^2} dt^2 - 
    (1+M/r)^2 (dr^2 + r^2 d\Omega^2)  \cr
   \tilde{F}_3 &= Q \vol_{S^2} \qquad G_3 = P \vol_{S^2}  \cr
   e^{2 \phi} &= 1 \qquad B = 0 \ .
  }}
 The electric charge $Q$, the magnetic charge $P$, and the mass $M$
are related by $Q = P = M/\sqrt{2}$.  By definition, $\vol_{S^2} =
\sin\theta \, d\theta \wedge d\phi$.  Gauss' law for the electric and
magnetic charges reads
  \eqn\GaussLaw{\vcenter{\openup2\jot
    \halign{\strut\span\TL & \span\TR \qquad & \span\TL & \span\TR\cr
   \int_{S^2} F_3 &= 0 & \int_{S^2} \tilde{F}_3 &= 4 \pi Q  \cr
   \int_{S^2} G_3 &= 4 \pi P & \int_{S^2} \tilde{G}_3 &= 0 \ .  \cr
  }}}
 Keeping the charges $Q$ and $P$ fixed but increasing the mass, one
obtains the non-extremal generalization of \kpSol:
  \eqn\NonExSol{\eqalign{
   ds^2 &= {h \over f^2} dt^2 - 
    f^2 \left( h^{-1} dr^2 + r^2 d\Omega^2 \right)  \cr
   \tilde{F}_3 &= Q \vol_{S^2} \qquad G_3 = P \vol_{S^2}  \cr
   e^{2 \phi} &= 1 \qquad B = 0 
  }}
 where
  \eqn\hfDef{
   h = 1 - {r_0 \over r} \qquad\qquad 
   f = 1 + {r_0 \sinh^2 \alpha \over r} \ .
  }
 The mass, charges, area, and temperature of this black hole are given
by
  \eqn\ThermQs{\vcenter{\openup1\jot
    \halign{\strut\span\TL & \span\TR \qquad\qquad & \span\TL & \span\TR\cr
   M &= {r_0 \over 2} \cosh 2\alpha &
   Q &= P = {r_0 \over 2 \sqrt{2}} \sinh 2\alpha  \cr
   A &= 4\pi r_0^2 \cosh^4 \alpha &
   T &= {1 \over 4 \pi r_0 \cosh^4 \alpha} \ .  \cr
  }}}
 Taking $\alpha \to \infty$ with $Q$ and $P$ held fixed, one recovers
the extremal solution \kpSol.

A crucial property of \kpSol\ and \NonExSol, without which there will
be no minimally coupled photons, is the vanishing of the dilaton.
This happens only when $Q=P$.  The situation is similar to the case of
fixed scalars, where the $Q=P$ case \cite{kr} was much easier to deal
with than the $Q \ne P$ case \cite{kkTwo}.  

When the equations \SEOMs\ are linearized around the solution
\NonExSol, the variations $\delta F_3^{\mu\nu}$ and $\delta
G_3^{\mu\nu}$ appear in all the equations of motion because the
background values of $F_3$ and $G_3$ are nonzero.  
But because $F_1$, $F_2$, $G_1$, and $G_2$ do have vanishing
background values (as does the axion $B$) the variations of these
gauge fields appear in the linearized equations only as
  \eqn\OnlyAppear{
   \nabla_\mu \big( e^{-2\phi} F^{\mu\nu} \big) = 0 \ ,
  }
 where $F = \delta F_1$, $\delta F_2$, $\delta G_1$, or $\delta G_2$.
Including $e^{-2\phi}$ in \OnlyAppear\ was unnecessary since it is
identically $1$ when the charges are equal; but \OnlyAppear\ is still
the right linearized equation of motion for these fields when the
charges are unequal (provided the background value of the axion
remains zero).  Surprisingly enough, a non-constant dilaton background
makes it much more difficult to decouple the equations for different
components of the gauge field.

\newsec{Separable equations for gauge fields}
\seclab\SepEq

Having shown that minimally coupled gauge fields do indeed exist, let
us now show how to solve their equations of motion.  The dilaton
background will be kept arbitrary just long enough to observe why it
makes the job much harder.  The goal of this section is to convert
Maxwell's equations
  \eqn\MaxwellEqs{
   d F = 0 \qquad d * e^{-2\phi} F = 0
  }
 into decoupled separable differential equations.  The equation of
motion for the vector potential $A$,
  \eqn\AEOM{
   d * e^{-2\phi} d A = 0 \ ,
  }
 does not lend itself to this task.  It turns out to be easier to
dispense with $A$ altogether and analyze Maxwell's equations,
\MaxwellEqs, directly.  Even this is quite challenging if one sticks
to the traditional tools of tensor analysis.  Fortunately, several
authors \cite{teuk,teukI,Churil} in the 60's and 70's worked out an
elegant approach to this sort of problem using Penrose's dyadic index
formalism \cite{np}.  Appendix~A provides a summary of some of the
standard notation.

The field strength $F_{\mu\nu}$ (six real quantities) is replaced by a
symmetric matrix $\Phi_{\Delta\Gamma}$ (three complex quantities)
using the equation
  \eqn\PhiDef{
   F_{\mu\nu} \sigma^\mu_{\Delta\dot\Delta} \sigma^\nu_{\Gamma\dot\Gamma} =
    \Phi_{\Delta\Gamma} \epsilon_{\dot\Delta\dot\Gamma} + 
    \bar\Phi_{\dot\Delta\dot\Gamma} \epsilon_{\Delta\Gamma} \ .
  }
 Now define $\phi_1$, $\phi_0$, and $\phi_{-1}$ as
follows:
  \eqn\MorePhiDefs{\eqalign{
   \phi_1 &= \Phi_{00} = F_{\mu\nu} \ell^\mu m^\nu \cr
   \phi_0 &= \Phi_{10} = \Phi_{01} 
     = \tf{1}{2} F_{\mu\nu} (\ell^\mu n^\nu + \bar{m}^\mu m^\nu) \cr
   \phi_{-1} &= \Phi_{11} = F_{\mu\nu} \bar{m}^\mu n^\nu  
  }}
 where $\ell^\mu$, $n^\mu$, $m^\mu$, and $\bar{m}^\mu$ form the
complex null tetrad (see the appendix).  In the literature, it is more
common to write $\phi_0$, $\phi_1$, and $\phi_2$ instead of $\phi_1$,
$\phi_0$, and $\phi_{-1}$.  The present convention has the advantage
that the subscript is essentially the helicity.

The Bianchi identity $d F = 0$ can be rewritten as
  $D^{\Gamma\dot\Delta} \Phi^\Delta{}_\Gamma = 
   D^{\Delta\dot\Gamma} \bar\Phi^{\dot\Delta}{}_{\dot\Gamma}$.
 Using this identity one can rewrite the equation of motion 
$d * e^{-2\phi} F = 0$ as
  \eqn\EOMReWrite{
   D^{\Gamma\dot\Delta} \Phi^\Delta{}_\Gamma = 
    \tf{1}{2} (\partial_{\Gamma\dot\Gamma} e^{-2\phi})
     (\Phi^{\Delta\Gamma} \epsilon^{\dot\Delta\dot\Gamma} + 
      \bar\Phi^{\dot\Delta\dot\Gamma} \epsilon^{\Delta\Gamma}) \ .
  }
 One immediately sees that the equations simplify greatly if the
coupling $e^{-2\phi} = 1$.  If this is not the case, then because the
right hand side involves $\bar\Phi_{\dot\Delta\dot\Gamma}$ as well as
$\Phi_{\Delta\Gamma}$, the advantage of compressing the real field
strength components into complex components of $\Phi_{\Delta\Gamma}$
is lost.  In this case, I have been unable to decouple the equations.
It is striking that the condition for Maxwell's equations to be simple
is the same as the condition found in \cite{kr,cgkt} for the fixed
scalar equation to decouple from Einstein's equations.
 
 Let us proceed with the case where $e^{-2\phi} = 1$, so
that Maxwell's equations can be succinctly written as
$D^{\Delta\dot\Delta} \Phi_{\Delta\Gamma} = 0$.  The spin coefficients
for the general spherically symmetric metric,
  \eqn\GenMetric{
   ds^2 = e^{2A(r)} dt^2 - e^{2B(r)} dr^2 - 
    e^{2C(r)} \left( d\theta^2 + \sin^2 \theta d\phi^2 \right) 
  }
 are presented in \DiagonalSC.  Six of them vanish, and the remaining
six can be expressed in terms of $\gamma$, $\rho$, and $\alpha$, which
are real.  Maxwell's equations written out in components therefore
take on a particularly simple form:
  \eqn\Max{\vcenter{\openup0\jot
    \halign{\strut\span\TL & \span\TR\cr
   (\Delta - 2 \gamma + \rho) \phi_1 &= \delta \phi_0 \cr
    (D - 2 \rho) \phi_0 &= (\bar\delta - 2 \alpha) \phi_1  
      \cr\noalign{\vskip1\jot}
   (\Delta + 2 \rho) \phi_0 &= (\delta - 2 \alpha) \phi_{-1} \cr
    (D + 2 \gamma - \rho) \phi_{-1} &= \bar\delta \phi_0  \ . \cr
  }}}
 The form of Maxwell's equations in a more general metric can be found
in \cite{np}.

A straightforward generalization of the preceding treatment can be
given for fields of arbitrary nonzero spin.  The simplest Lorentz
covariant wave equation for a massless field of spin $n/2$ is
  \eqn\SimpleEOM{
    D^{\Delta_1 \dot\Delta} \Psi_{\Delta_1 \ldots \Delta_n} = 0
  }
 where $\Psi_{\Delta_1 \ldots \Delta_n}$ is symmetric in all its
indices.  The case $n=1$ gives the Weyl fermion equation.  The case
$n=2$ is, as we have seen, Maxwell's equations in vacuum.  The case
$n=4$ can be obtained by linearizing pure gravity around Minkowski
space, as discussed in section 5.7 of \cite{pr}.  The case $n=3$ can
be obtained in Minkowski space from the massless Rarita-Schwinger
equation, as follows.  The constraint $\gamma^\mu \psi_\mu = 0$ is
imposed on the Rarita-Schwinger field
  \eqn\RSPot{
   \psi^\mu = 
    \pmatrix{
     \sigma^{\mu\Delta\dot\Delta} \psi_{\Delta\Gamma\dot\Delta}
       \cr\noalign{\vskip5pt}
     \sigma^{\mu\dot\Delta\Delta} 
      \bar\psi_{\dot\Delta}{}^{\dot\Gamma}{}_{\vphantom{\dot\Delta}\Delta}
    }
  }
 to project out the spin-$1/2$ components.  This constraint is
equivalent to making $\psi_{\Delta\Gamma\dot\Delta}$ symmetric in its
two undotted indices.  In the supergravity literature, the equation of
motion is usually written as $\epsilon^{\mu\nu\rho\sigma} \gamma_5
\gamma_\nu \Psi_{\rho\sigma} = 0$ where $\Psi_{\rho\sigma} =
\partial_\rho \psi_\sigma - \partial_\sigma \psi_\rho$.  The original
paper by Rarita and Schwinger \cite{rs} (see also p.~323 of
\cite{weinbergI}) proposes $\slashed\partial \psi_\mu = 0$ as the
equation of motion.  Using the constraint one can show that both are
equivalent to $\partial^{\dot\Delta\Delta}
\psi_{\Delta\Gamma\dot\Sigma} = 0$.  As a result, the field strength
  \eqn\RSField{
   \Psi_{\Delta\Sigma\Gamma} = \partial_{\Sigma\dot\Delta}
    \psi_\Delta{}^{\dot\Delta}{}_\Gamma 
  }
 is symmetric in all its indices and obeys the $n=3$ case of
\SimpleEOM.  

Although the cases $n=3$ and $n=4$ of \SimpleEOM\ are not in general
the correct curved-space equations of motion for the gravitino and
graviton, and although for $n>2$ there are problems defining local,
gauge-invariant number currents and stress-energy tensors, still a
brief investigation of \SimpleEOM\ serves to illustrate some of the
general features one expects for fields of higher spin.  Furthermore,
the near-Minkowskian limit of the equations I will derive should be
close in form to the actual graviton and gravitino equations far from
a black hole.

Define helicity components $\psi_s$ according to
  \eqn\HComp{
   \Psi_{\Delta_1 \ldots \Delta_n} =
    \psi_{\lower4pt\hbox{$\scriptstyle {{n \over 2} - 
     \sum\limits_i \Delta_i}$}}  \ .
  }
\vskip-5pt
 Then \SimpleEOM\ can be written out in components.  There are $2n$
equations:
  \eqn\AllEOMS{\vcenter{\openup0\jot
    \halign{\strut\span\TL & \span\TR\cr
     \big( \Delta - n \gamma + \rho \big) \psi_{n \over 2} &=
      \big( \delta + (n-2) \alpha \big) \psi_{{n \over 2} - 1}  \cr
     \big( D - (n-2) \gamma - n \rho \big) \psi_{{n \over 2} - 1} &=
      \big( \bar\delta - n \alpha \big) \psi_{n \over 2}  \cr
     &\vdots  \cr
     \big( \Delta - 2s \gamma + (\tf{n}{2} + 1 - s) \rho \big) \psi_s &=
      \big( \delta + (2s - 2) \alpha \big) \psi_{s-1}  \cr
     \big( D - (2s - 2) \gamma - (\tf{n}{2} + s) \rho \big) \psi_{s-1} &=
      \big( \bar\delta - 2s \alpha \big) \psi_s  \cr
     &\vdots  \cr
     \big( \Delta + (n-2) \gamma + n \rho \big) \psi_{-{n \over 2} + 1} &=
      \big( \delta - n \alpha \big) \psi_{-{n \over 2}}  \cr
     \big( D + n \gamma - \rho \big) \psi_{-{n \over 2}} &= 
      \big( \bar\delta + (n-2) \alpha \big) \psi_{-{n \over 2} + 1} 
        \ .  \cr
   }}} 
 The equations \AllEOMS\ are invariant under PT, which sends $\psi_s
\to \psi_{-s}$, $\gamma \to -\gamma$, $\rho \to -\rho$, $D
\leftrightarrow \Delta$, $\delta \leftrightarrow \bar\delta$.

The commutation relations
  \eqn\KeyComs{
   [D,\delta] = \rho \delta \ \ \quad [D,\alpha] = \rho \alpha \ \ \quad
   [\Delta,\bar\delta] = -\rho \bar\delta \ \ \quad
   [\Delta,\alpha] = -\rho \alpha
  }
 are easily established by direct computation.  They can be used to
convert the pair of equations in \AllEOMS\ relating $\psi_s$ and
$\psi_{s-1}$ into decoupled second order equations for $\psi_s$ and
$\psi_{s-1}$ separately.  In this way one obtains
  \eqn\SameEQ{\eqalign{
   &\Big[ \big( D - (2s-2) \gamma - (\tf{n}{2}+1+s) \rho \big)
           \big( \Delta - 2s \gamma + (\tf{n}{2}+1-s) \rho \big)  \cr
   &\qquad - \big( \delta + (2s-2) \alpha \big) 
           \big( \bar\delta - 2s \alpha \big) \Big] \psi_s = 0  \cr
   &\Big[ \big( \Delta - (2s+2) \gamma + (\tf{n}{2}+1-s) \rho \big)
           \big( D - 2s \gamma - (\tf{n}{2}+1+s) \rho \big)  \cr
   &\qquad - \big( \bar\delta - (2s+2) \alpha \big)
           \big( \delta + 2s \alpha \big) \Big] \psi_s = 0 \ .
  }}
 The first of these can be derived for $s > -n/2$, while the second
can be derived for $s < n/2$.  In fact they are different forms of the
same equation, which can be written out more simply in terms of the
fields 
  \eqn\TPsi{
   \tilde\psi_s = 
    e^{|s| A + \left( {n \over 2} - |s| + 1 \right) C} \psi_s 
  }
 as
  \eqn\TPsiEqs{\vcenter{\openup1\jot
    \halign{\strut\span\TL & \span\TR & \span\TT\cr
     \Big[ \big( D + (2-4s) \gamma - 2s \rho \big) \Delta - 
           \big( \delta + (2s-2) \alpha \big) 
            \big( \bar\delta - 2s \alpha \big) \Big] \tilde\psi_s &= 0
      &\quad\hbox{for $s \ge 0$}  \cr
     \Big[ \big( \Delta - (2+4s) \gamma - 2s \rho \big) D - 
           \big( \bar\delta - (2s+2) \alpha \big)
            \big( \delta + 2s \alpha \big) \Big] \tilde\psi_s &= 0
      &\quad\hbox{for $s \le 0$.}  \cr
  }}}
 More explicitly, 
  \eqn\OneMaster{\eqalign{
   &\Big[ \partial_r^2 + 
          \big( (1-2|s|) A' - B' + 2 |s| C' \big) \partial_r - 
          e^{-2A+2B} \partial_t^2 +
          2s \, e^{-A+B} (A'-C') \partial_t  \cr
   &\qquad + e^{2B-2C} \big( 
           \partial_\theta^2 + \cot\theta \partial_\theta + 
           \csc^2 \theta \partial_\phi^2 + 
           2is \cot\theta \csc\theta \partial_\phi - 
           s^2 \cot^2 \theta - |s| \big)
    \Big] \tilde\psi_s = 0 
  }}
 for all values of $s$, positive and negative.  It is interesting to
note that there is no explicit dependence on the spin $n/2$ of the
particle in \OneMaster, only on its helicity $s$.  In practice we will
mainly be interested in the equations for $\tilde\psi_{\pm {n \over
2}}$ since these are the only components that can be radiative.  The
fact that the equations for these radiative fields are identical to
equations obeyed by non-radiative components of fields of higher spin
suggests that mixing of different spins is possible.  Such mixing
between photons and gravitons was observed by Chandrasekhar in his
analysis of perturbations of the Reissner-Nordstrom black hole
\cite{Chandra}.

Equations similar to \OneMaster\ were worked out for the Kerr metric
by Teukolsky \cite{teuk,teukI}.  In that case, only the equations
for the radiative fields turned out to be separable.  But in the
present context, spherical symmetry makes the separability of
\OneMaster\ trivial: the general solution is
  \eqn\SepSol{
   \tilde\psi_s(t,r,\theta,\phi) = 
    e^{-i \omega t} R_{s\ell}(r) Y_{s\ell m}(\theta,\phi)
  }
 where $Y_{s\ell m}$ is a spin-weighted spherical harmonic \cite{gold}
and $R_{s\ell}(r)$ satisfies the ODE
  \eqn\RadialEquation{\eqalign{
   &\Big[ \partial_r^2 + 
          \big( (1-2|s|) A' - B' + 2|s| C' \big) \partial_r + 
          \omega^2 e^{-2 A + 2 B} - 
          2si \omega \, e^{-A + B} (A' - C')   \cr
   &\qquad - e^{2 B - 2 C} (\ell+|s|) (\ell-|s|+1) 
    \Big] R_{s\ell} = 0 \ .
  }}
 The minimal value of $\ell$ is $|s|$.  In the case of photons, $\ell
\ge 1$ indicates that the fields of lowest moment that can be radiated
are dipole fields.  For fermions, $\ell$, $s$, and $m$ are all
half-integer.

\newsec{Semi-classical absorption probabilities}
\seclab\Prob

Returning now to the case of photons, let us investigate how an
absorption cross-section can be extracted from a solution to
\RadialEquation.  Section~\Prob.\Method\ derives a formula for the
absorption probability.  In section~\Prob.\Examples\ matching solutions
are exhibited and absorption probabilities calculated for the black
holes \kpSol\ and \NonExSol.

\subsec{Probabilities from energy fluxes}
\subseclab\Method

For the photon as for other fields of spin greater
than $1/2$, there is no gauge invariant number current analogous to
  $J_\mu = {1 \over 2 i} \bar\phi 
    \overleftrightarrow\partial_\mu \phi$
 for spin $0$ and 
  $J_\mu = \bar\psi \gamma_\mu \psi$
 for spin $1/2$.  In order to count the photons falling into the black
hole, it is therefore necessary to examine the energy flux through the
horizon and adjust for the gravitational blueshift that the infalling
photons experience.  The stress-energy tensor can be written in terms
of $\phi_1$, $\phi_0$, and $\phi_{-1}$:
  \eqn\Tmunu{\eqalign{
   T^{\mu\nu} &= \tf{1}{4} g^{\mu\nu} F^2 + F^{\mu\rho} F_\rho{}^\nu  
     = 2 \sigma^{\mu\Delta\dot\Delta} \sigma^{\nu\Gamma\dot\Gamma}
     \phi_{\Delta\Gamma} \bar\phi_{\dot\Delta\dot\Gamma}  \cr
    &= \Big[ |\phi_1|^2 n_\mu n_\nu + 
     2 |\phi_0|^2 (\ell_{(\mu} n_{\nu)} + m_{(\mu} \bar{m}_{\nu)}) + 
     |\phi_{-1}|^2 \ell_\mu \ell_\nu  \cr
    &\quad\ -4 \bar\phi_1 \phi_0 n_{(\mu} m_{\nu)} - 
     4 \bar\phi_0 \phi_{-1} \ell_{(\mu} m_{\nu)} + 
     2 \phi_{-1} \bar\phi_1 m_\mu m_\nu \Big] + {\rm c.c.}
  }}
 Consider a sphere $S^2$ located anywhere outside the horizon.
Taking into account the blueshift factor as described, the number of
photons passing through $S^2$ in a time interval $[0,t]$ is 
  \eqn\NumberThrough{
   N = {1 \over \omega} \int\limits_{S^2 \times [0,t]} * (T_{tr} dr) 
     = {t \over \omega} \int\limits_{S^2} {\cal F} \vol_{S^2}
  }
 where $\vol_{S^2} = \sin\theta d\theta \wedge d\phi$, and 
  \eqn\FFlux{
   {\cal F} = e^{A-B+2C} T_{tr} 
    = e^{2 A + 2 C} \left( |\phi_{-1}|^2 - |\phi_1|^2 \right)
    = |\tilde\phi_{-1}|^2 - |\tilde\phi_1|^2
  }
 is essentially the radial photon number flux.

The goal now is to find an approximate solution to \OneMaster\ for
photons whose wavelength is much longer than the size of the black
hole, and to extract from it an absorption probability and
cross-section.  The dominant contribution to this absorption comes
from dipole fields.  

Far from the black hole, \RadialEquation\ for dipole fields
simplifies to
  \eqn\FarPRE{
   \left[ \partial_\rho^2 + {2 \over \rho} \partial_\rho + 1 + 
    {2 s i \over \rho} - {2 \over \rho^2} \right] R = 0
  }
 where $\rho = \omega r$.  The general solution to \FarPRE\ with $s =
1$ is 
  \eqn\FarSoln{
   R = 2 a e^{-i \rho} \left( 1 - {i \over \rho} - 
       {1 \over 2 \rho^2} \right) + b {e^{i \rho} \over \rho^2} \ .
  }
 The general solution with $s = -1$ is just the conjugate of \FarSoln.
By using \Max\ $\phi_0$ can be calculated as well.  Let us choose the
spatial orientation by setting $m=0$ in \SepSol.  Then the final
result is
  \eqn\IIISoln{\eqalign{
   \tilde\phi_1 &= e^{-i \omega t} \sin\theta
    \left[ 2 a e^{-i \rho} \left( 1 - {i \over \rho} - 
      {1 \over 2 \rho^2} \right) + b {e^{i \rho} \over \rho^2}
    \right]  \cr
   \tilde\phi_0 &= e^{-i \omega t} \cos\theta {\sqrt{2} i \over \omega}
    \left[ a e^{-i \rho} \left( 1 - {i \over \rho} \right) + 
           b e^{-i \rho} \left( 1 + {i \over \rho} \right) 
    \right]  \cr
   \tilde\phi_{-1} &= e^{-i \omega t} \sin\theta 
    \left[ a {e^{-i \rho} \over \rho^2} + 2 b e^{i \rho} \left( 1 +
     {i \over \rho} - {1 \over 2 \rho^2} \right) \right] \ .
  }}
 It is clear from \IIISoln\ that $\phi_{-1}$ is the radiative component
of the field for outgoing waves, while $\phi_1$ is the radiative
component for ingoing waves.  

The boundary conditions at the horizon \cite{teukI} require the radial
group velocity to point inward.  It can be shown that $\tilde\phi_1$
remains finite at the horizon while $\tilde\phi_{-1}$ vanishes.
Normalizations are fixed by requiring $|\tilde\phi_1|^2 \to \sin^2
\theta$ at the horizon.  The net flux of photons into the black hole
can be computed in two ways:
  \eqn\FluxesFN{\vcenter{\openup1\jot
   \halign{\strut\span\TT & \span\TR\cr
    at the horizon: \quad & {\cal F}_h = -|\tilde\phi_1|^2 = 
     -\sin^2 \theta  \cr
    at infinity: \quad & {\cal F}_\infty = 
     {\cal F}^{\rm out}_\infty + {\cal F}^{\rm in}_\infty = 
     |\tilde\phi_{-1}|^2 - |\tilde\phi_1|^2 = 
      4 \left( |b|^2 - |a|^2 \right) \sin^2 \theta \ .  \cr
  }}}
 The two must agree, ${\cal F}_h = {\cal F}_\infty$, and so $|a|^2 =
|b|^2 + 1/4$.  One can thus easily perceive the equivalence of the two
common methods for computing the absorption probability.  The first
examines the deficit in the outgoing flux compared to the ingoing
flux:
  \eqn\Deficit{
   1 - P = {{\cal F}^{\rm out}_\infty \over {\cal F}^{\rm in}_\infty} 
         = {|b|^2 \over |a|^2} \ ,
  }
 while the second simply compares the flux on the horizon to the
ingoing flux at infinity:
  \eqn\FluxRatio{
   P = {{\cal F}_h \over {\cal F}^{\rm in}_\infty} 
     = {1 \over 4 |a|^2} \ .
  }
 For low-energy photons, the absorption probability is small and $a$
and $b$ are large and nearly equal, so from a calculational point of
view the second method is to be preferred over the first.  Indeed, it
will be standard practice in the matching calculations of later
sections to ignore the small difference between $a$ and $b$ and simply
set them equal.  This approximation suffices when \FluxRatio\ is used.
 
Finally, to obtain the absorption cross-section from the probability,
the Optical Theorem is needed.  Averaging over polarizations is
unnecessary in view of the spherical symmetry of the background.  The
result for photons in a dipole wave is 
  \eqn\OpThPh{
   \sigma_{\rm abs} = {3 \pi \over \omega^2} P \ .
  }

\subsec{Matching solutions}
\subseclab\Examples

The work of previous sections can be boiled down to a simple
prescription for computing the absorption probability and
cross-section to leading order in the energy for minimally coupled
photons falling into a spherically symmetric black hole.  The
probability can be obtained by solving \RadialEquation\ with $\ell =
1$ and $s = 1$, subject to the boundary condition $R(r) \sim e^{i
f(r)}$ as $r$ approaches the horizon, $f(r)$ being some real
decreasing function of $r$.  Far from the black hole, one will find
$R(r) \sim 2 a e^{-i \omega r}$, and the probability is then given by
  \eqn\ReState{\eqalign{
   P = {1 \over 4 |a|^2} 
     = {|R(r)|^2 \Big|_{\rm horizon} \over 
        |R(r)|^2 \Big|_\infty} \ .
  }}

First consider the extreme black hole \kpSol.  The radial equation
\RadialEquation\ with $\ell = s = 1$ is
  \eqn\ExactExtremeI{
   \left[ \partial_r^2 + {2 \over r + M} \partial_r + 
    \omega^2 \left( 1 + {M \over r} \right)^4 + 
    {2 i \omega \over r} \left( 1 - {M^2 \over r^2} \right) - 
    {2 \over r^2} \right] R = 0 \ .
  }
 A different radial variable, $y = \omega M^2 / r$, is more natural
near the horizon.  In terms of $y$, \ExactExtremeI\ can be rewritten
as
  \eqn\ExactExtremeII{
   \left[ (y^2 \partial_y)^2 - 
    {2 \omega M \over y + \omega M} y^3 \partial_y + 
    (y+\omega M)^4 - 2 i y (y^2 - \omega^2 M^2) - 
    2 y^2 \right] R = 0 \ .
  }
 A matching solution can be pieced together as usual from a near
region (${\bf I}$), an intermediate region (${\bf II}$), and a far
region (${\bf III}$).  In the near region, \ExactExtremeII\ is
simplified by setting to zero all terms containing explicit factors of
$\omega M$.  The intermediate region solution is obtained from
\ExactExtremeI\ with $\omega = 0$.  In the far region, we simply make
the flat space approximation, obtaining \FarPRE\ and \FarSoln.  
The solutions in the three regions are
  \eqn\OTTSoln{\eqalign{
   R_{\bf I} &= e^{i y} \left( 1 + {i \over y} - 
    {1 \over 2 y^2} \right)  \cr
   R_{\bf II} &= {C_1 r \over 1+M/r} + {C_2 \over r^2 (1+M/r)}  \cr
   R_{\bf III} &= 2 a e^{-i \rho} \left( 1 - {i \over \rho} - 
    {1 \over 2 \rho^2} \right) + b {e^{i \rho} \over \rho^2} \ .
  }}
 A match is obtained by setting
  \eqn\MatchIt{\eqalign{
   &C_1 = -{1 \over 2 \omega^2 M^3} \qquad C_2 = 0  \cr
   &a = b = -{3 i \over 8 (\omega M)^3} \ .
  }}
 The absorption probability and cross-section are 
  \eqn\PandSigma{
   P = {16 \over 9} (\omega M)^6 \qquad\quad 
   \sigma_{\rm abs} = {16 \pi \over 3} \omega^4 M^6 \ .
  }

Now consider the non-extremal generalization, \NonExSol.  The radial
equation \RadialEquation\ with $\ell = 1$ and $s = 1$ is
  \eqn\ExNonEx{
   \left[ \partial_r^2 + {2 \over fr} \partial_r + 
    \omega^2 {f^4 \over h^2} - 2 i \omega \left( \tf{1}{2} + 
    {1 \over 2h} - {2 \over f} \right) - {2 \over hr^2} \right] R = 0 \ .
  }
 As before, the far region is treated in the flat space approximation,
and a solution in the intermediate region is obtained by solving
\ExNonEx\ with $\omega = 0$.  In the near region, the useful radial
variable is $h$ itself.  Having the black hole near extremality is
useful since one can approximate $f \approx (1-h) \cosh^2 \alpha$ and
drop terms in \ExNonEx\ which are small in the limit where $\alpha \to
\infty$ and $\omega r_0 \to 0$ with
  \eqn\lambdaDef{
   \lambda = \omega r_0 \cosh^4 \alpha = {\omega \over 4 \pi T}
  }
 held fixed.  The result is that \ExNonEx\ simplifies to
  \eqn\INonEx{
   \left[ h (1-h) \partial_h^2 - 2h \partial_h + 
    \lambda^2 {1-h \over h} - i\lambda {1+h \over h} - 
    {2 \over 1-h} \right] R = 0 \ ,
  }
 which is representative of the general form of differential equation
which is solved by a hypergeometric function of $h$ times powers of
$h$ and $1-h$.

The solutions in the three regions are 
  \eqn\NonExSols{\eqalign{
   R_{\bf I} &= {h^{-i \lambda} \over (1-h)^2} 
     F(-2,-1-2i\lambda,-2i\lambda;h)  \cr
    &= {h^{-i \lambda} \over (1-h)^2} \left( 1 + 
     {ih (1+2i\lambda) \over \lambda} - 
     {h^2 (1+2i\lambda) \over 1-2i\lambda} \right)  \cr
   R_{\bf II} &= C_1 {h \over f (1-h)} + 
    C_2 {h \over f} \left( 1 + {1 \over h} + 
    {2 \log h \over 1-h} \right)  \cr
   R_{\bf III} &= 2 a e^{-i \rho} \left( 1 - {i \over \rho} - 
    {1 \over 2 \rho^2} \right) + b {e^{i \rho} \over \rho^2} \ ,
  }}
 and a match is obtained by setting
  \eqn\MatchNonEx{\eqalign{
   &C_1 = {i \cosh^2 \alpha \over \lambda (1-2i\lambda)} \qquad 
    C_2 = 0  \cr
   &a = b = -{3 \cosh^2 \alpha \over 4 \omega r_0 \lambda
    (1-2i\lambda)} \ .
  }}
 The absorption probability and cross-section are 
  \eqn\NonExProb{
   P = \tf{4}{9} (\omega r_0)^3 \lambda (1+4\lambda^2) \qquad\quad 
   \sigma_{\rm abs} = {4 \pi \over 3} 
    \omega r_0^3 \lambda (1+4\lambda^2) \ .
  }

\newsec{The axion and minimally coupled fermions}
\seclab\OtherParticles

Because the solution \kpSol\ preserves a quarter of the supersymmetry,
it is clearly of interest to compare cross-sections of particles with
different spins related by the unbroken supersymmetry.  The other
particles in $N=4$ supergravity whose cross-sections are
straightforward to compute are the dilaton, the axion, and those
fermions which obey the Weyl equation.  The dilaton has been dealt
with at length in the fixed scalar literature \cite{kr,kkOne,kkTwo}.
The axion in fact is also a fixed scalar, as section
\OtherParticles.\Axion\ will show.  Of the four massless fermions, two
are minimally coupled.  The purpose of section
\OtherParticles.\Fermions\ is to demonstrate this fact and to compute
the minimal fermion cross-section.  For comparison with \NonExProb\ I
will quote here the final results:
  \eqn\AFProbs{\vcenter{\openup1\jot
    \halign{\strut\span\TT\ \ & \span\TL & \span\TR \qquad & 
            \span\TL & \span\TR\cr
     axion, dilaton: &    
      P &= (\omega r_0)^2 (1+4\lambda^2) &
      \sigma_{\rm abs} &= \pi r_0^2 (1+4\lambda^2)  \cr
     minimal fermions: &
      P &= {(\omega r_0)^2 \over 4} (1 + 16 \lambda^2) &
      \sigma_{\rm abs} &= {\pi r_0^2 \over 2} (1 + 16 \lambda^2)  \cr
  }}}
 where $\lambda = \omega / (4 \pi T)$, as in \NonExProb.  Note that in
the extremal limit the bosonic and fermionic absorption probabilities
quoted in \AFProbs\ coincide.\foot{Similar results on the agreement of
absorption probabilities due to residual supersymmetry have appeared
elsewhere in the literature \cite{OkaOne,OkaTwo} for the case of $N=2$
supergravity.  See also \cite{krw}.  Thanks to G.~Horowitz and A.~Peet
for bringing these papers to my attention.}

\subsec{The axion}
\subseclab\Axion

The strategy for deriving the linearized equation of motion for the
axion is the same as the one used in \cite{kr} for the dilaton: only
spherical perturbations of the solution \NonExSol\ are considered, and
a gauge is chosen where the only components of the metric that
fluctuate are $g_{rr}$ and $g_{tt}$.  The minimally coupled gauge
fields $F_1$, $F_2$, $G_1$, and $G_2$ do not affect the linearized
axion equation because they have no background value and enter into
the axion equation quadratically.  Spherical symmetry dictates that
only $tr$ and $\theta\phi$ components of these field strengths can
fluctuate, corresponding respectively to radial electric and radial
magnetic field fluctuations.  These fluctuations are constrained
further by Gauss' law \GaussLaw:
  \eqn\GLConsequence{\vcenter{\openup1\jot
    \halign{\strut\span\TL & \span\TR \qquad & \span\TL & \span\TR\cr
   F^3_{\theta\phi} &= 0 & 
   F_3^{tr} &= {Q \sin\theta \over \sqrt{-g}} e^{2\phi}  \cr
   G^3_{\theta\phi} &= P \sin\theta &
   G_3^{tr} &= {2 B e^{2\phi} \over \sqrt{-g}} G^3_{\theta\phi} \ .  \cr
  }}}
 The asymmetry between $F_3$ and $G_3$ arises because the black hole
is electrically charged under $F_3$ and magnetically charged under
$G_3$.  The axion $B$ is a dynamical theta-angle for the gauge fields,
so the last relation in \GLConsequence\ should be viewed as a
dynamical version of the Witten effect: when the axion fluctuates, an
object that was magnetically charged picks up what seems like an
electric charge in that there are radial electric fields.

Using \GLConsequence\ and ignoring the dilaton terms in the axion
equation of \SEOMs\ (which is valid for the purpose of deriving the
linearized axion equation because the dilaton has zero background
value), one obtains
  \eqn\SimpleAxion{
   \square B + \tf{i}{2} (F_3 * F_3 + G_3 * G_3) = 
   \square B + i (G_3^{tr} * G^3_{tr} + G_3^{\theta\phi} * G^3_{\theta\phi}) =
   \left[ \square + {4 P^2 \over f^4 r^4} \right] B = 0 \ .
  }
 This is indeed identical to the linearized equation for the dilaton,
although the ``mass'' term for the dilaton receives equal
contributions $2 Q^2 + 2 P^2$ from the electric and magnetic charges
in place of the $4 P^2$ we see in \SimpleAxion.

The radial equation for $B$ and $\phi$ is 
  \eqn\DilRad{
   \left[ (hr^2 \partial_r)^2 + \omega^2 r^4 f^4 - 
    {h r_0^2 \sinh^2 2\alpha \over 2 f^2} \right] R = 0  \ .
  }
 By an analysis sufficiently analogous to the treatments in
\cite{kr,kkOne} that it seems superfluous to present the details, one
obtains the result already quoted in \AFProbs:
  \eqn\DilProb{
   P = (\omega r_0)^2 (1+4\lambda^2) \qquad\quad 
   \sigma_{\rm abs} = \pi r_0^2 (1+4\lambda^2) \ .
  }

\subsec{Minimally coupled fermions}
\subseclab\Fermions

It was shown in \cite{csfTwo} that the complete fermionic equations of
motion for simple $N=4$, $d=4$ supergravity take on a simple form when
written in terms of the supercovariant derivatives introduced in
\cite{cs}.  The relevant one of these equations for spin-$1/2$
fermions is
  \eqn\FEOM{
   i \hat{\slashed{D}} \Lambda_I - \tf{3}{2} e^{2 \phi} 
    (\hat{\slashed{D}} B) \Lambda_I = 0 \ , 
  }
 where $\hat{D}_\mu$ denotes a supercovariant derivative.\foot{The
spinor and gamma matrix conventions conventions used here are those
described in Appendix~A of \cite{klopp}.  $\Lambda_I$ is a chiral
spinor with $\gamma_5 \Lambda_I = \Lambda_I$ which replaces the
Majorana spinor $\chi^i$ of \cite{csfTwo}.  The gravitinos are also
written in terms of chiral spinors $\Psi_\mu^I$ with $\gamma_5
\Psi_\mu^I = \Psi_\mu^I$.  $I$ runs from $1$ to $4$.  Conversion to
these conventions from those of \cite{csfTwo} is discussed in
\cite{klopp}.  I would only add that in the current conventions, each
field is identified with $K$ times its counterpart in \cite{csfTwo},
and for notational simplicity $K$ is then set equal to $1/2$.  With
this choice of the gravitational constant, $\phi$ and $\Lambda_I$ are
not canonically normalized; rather, they are twice the canonically
normalized fields.  So for example the kinetic term of $\phi$ in
\FullBL\ is $2 (\partial_\mu \phi)^2$ rather than the canonical ${1
\over 2} (\partial_\mu \phi)^2$.}

Supercovariant derivatives in general can be read off from the
supersymmetry variations of a field: if $\delta f = F_I \epsilon^I$,
then $\hat{D}_\mu f = D_\mu f - \tf{1}{4} F_I \Psi_\mu^I$.  Thus the
supercovariant derivative of the axion is $\hat{D}_\mu B =
\partial_\mu B + \hbox{(two fermion terms)}$.  Again, terms in the
equations of motion which are quadratic in fields with zero background
value do not contribute to the linearized first-varied equations of
motion.  Because the background values of the axion as well as all
fermions are zero for the solution \NonExSol, the second term in
\FEOM\ can be discarded.

A further simplification of \FEOM\ can be made by dropping terms from
$\hat{\slashed{D}} \Lambda_I$ which are quadratic in fields with zero
background value.  The supersymmetry variation of $\Lambda_I$ is
  \eqn\SUSYvs{
   \delta \Lambda_I = \sqrt{2} \sigma^{\rho\sigma} 
    (F^3_{\rho\sigma} \alpha^3_{IJ} - \tilde{G}^3_{\rho\sigma}
     \beta^3_{IJ}) \epsilon^J 
  }
 plus terms which vanish for the solution \NonExSol.  So
  \eqn\LambdaD{
   \hat{D}_\mu \Lambda_I = D_\mu \Lambda_I - 
    {1 \over 2 \sqrt{2}} \sigma^{\rho\sigma} 
    (F^3_{\rho\sigma} \alpha^3_{IJ} - \tilde{G}^3_{\rho\sigma}
     \beta^3_{IJ}) \Psi_\mu^J 
  }
 plus terms quadratic in fields with zero background value.  In
\SUSYvs\ and \LambdaD, I have introduced 
  $\sigma^{\rho\sigma} = \tf{1}{4} [\gamma^\rho,\gamma^\sigma]$ 
 and the matrices
  \eqn\ABDef{
   \alpha^3_{IJ} = \pmatrix{ 0 & 1 & 0 & 0 \cr
                            -1 & 0 & 0 & 0 \cr
                             0 & 0 & 0 & 1 \cr
                             0 & 0 &-1 & 0 } \qquad\quad
    \beta^3_{IJ} = \pmatrix{ 0 &-1 & 0 & 0 \cr
                             1 & 0 & 0 & 0 \cr
                             0 & 0 & 0 & 1 \cr
                             0 & 0 &-1 & 0 } \ .
  }
 The matrices $\alpha^n_{IJ}$ and $\beta^n_{IJ}$ were introduced in
\cite{gso} and used in \cite{csfTwo} to establish the $SU(4)$
invariance of $N=4$ supergravity.

Since $F^3_{\rho\sigma} = \tilde{G}^3_{\rho\sigma}$, the first
variation of the equation of motion \FEOM\ is
  \eqn\FirstFEOM{
   i \hat{\slashed{D}} \delta \Lambda_I = 
    i \slashed{D} \delta \Lambda_I - {i \over 2 \sqrt{2}} \sigma^{\rho\sigma}
    F^3_{\rho\sigma} (\alpha^3_{IJ} - \beta^3_{IJ}) 
     \delta \Psi_\mu^J = 0 \ .
  }
 For $I=3$ and $4$, the gravitino part gets killed and \FEOM\ is
nothing but the Weyl equation \WeylEq.  For $I=1$ and $2$, the
gravitino mixes in.\foot{The gravitino equation of motion has the form
of the Rarita-Schwinger equation plus interactions.  This equation
also has an $SO(4)$ index structure which is block diagonal, and the
spin-$1/2$ particles decouple from the $I = 3,4$ equations.  The
resulting gravitino equation,
  $\epsilon^{\mu\nu\rho\sigma} \gamma_\nu \hat{\Psi}^I_{\rho\sigma} = 0$,
 is less simple than the equations studied previously because the
supercovariant field strength $\hat{\Psi}^I_{\rho\sigma}$ involves a
non-vanishing combination of the field strengths $F_3$ and
$\tilde{G}_3$.  The problem of extracting from it a separable PDE like
\OneMaster\ is under investigation.}

The key point in this analysis is that the supersymmetry
transformation \SUSYvs\ leaves two of the $\Lambda_I$ invariant no
matter what $\epsilon^J$ is chosen to be.  In view of the form of
\FEOM\ and the prescription for reading off supercovariant derivatives
from the supersymmetry transformation laws, this makes it inevitable
that two of the $\Lambda_I$ are minimally coupled fermions.  The
situation is related to unbroken supersymmetries, but only loosely:
the non-extremal black hole has the same minimally coupled fermions
that the extremal one does, and the extremal black hole preserves only
one supersymmetry but admits two minimally coupled fermions.  The
condition for an unbroken supersymmetry is that the supersymmetry
variations of all the fermions fields must vanish.  So one would
expect that there are at least as many minimally coupled fermions as
unbroken supersymmetries.  More precisely, suppose that there are $n$
{\em broken} supersymmetries and $m$ fermions in the theory whose
equation of motion is of the form $i \hat{\slashed{D}} \Lambda = 0$,
up to terms quadratic in fields with zero background value.  Then
there must be at least $m-n$ minimally coupled fermions.  For
axion-free solutions to pure $N=4$ supergravity, $m = 4$.

The existence of minimally coupled fermions having been established,
the computation of their absorption cross-section now proceeds in
parallel to the case of minimally coupled photons.  If $R(r)$ is a
solution of the radial equation \RadialEquation\ with $\ell = s =
1/2$, then the absorption probability is once again
  \eqn\ProbAgain{
   P = {|R(r)|^2 \Big|_{\rm horizon} \over 
        |R(r)|^2 \Big|_\infty} \ .
  }
 The justification for \ProbAgain\ is a little different than for its
analog \ReState\ for photons because for fermions there is a conserved
number current, 
  \eqn\JDef{
   J_\mu = -\sqrt{2} \sigma_\mu^{\dot\Delta\Delta} \bar\Psi_{\dot\Delta} 
    \Psi_{\vphantom{\dot\Delta}\Delta} \ .
  }
 Here $\Psi_\Delta$ is the dyadic version of $\delta \Lambda_3$ or
$\delta \Lambda_4$.  The number of fermions passing through a sphere in
a time $t$ is
  \eqn\NPass{
   N = \int\limits_{S^2 \times [0,t]} * (J_r dr) 
     = t \int\limits_{S^2} {\cal F} \vol_{S^2}
  }
 where
  \eqn\FPass{
   {\cal F} = e^{A-B+2C} J_r 
    = e^{A+2C}
       \left( |\psi_{-1/2}|^2 - |\psi_{1/2}|^2 \right)
    = |\tilde\psi_{-1/2}|^2 - |\tilde\psi_{1/2}|^2 \ .
  }
 The component $\psi_{1/2}$, like $\phi_1$ in the case of photons, is
both the radiative component at infinity for infalling solutions and
the nonzero component at the black hole horizon.  The formula
\ProbAgain\ thus follows from \FPass\ by the same analysis that gave
\ReState\ from \FFlux.

The radial equation \RadialEquation\ with $\ell = s = 1/2$ is 
  \eqn\FermionRE{
   \left[ \partial_r^2 + {1 \over r} 
    \left( \tf{1}{2} + {1 \over 2 h} \right) + 
    \omega^2 {f^4 \over h^2} - i \omega {f^2 \over h} {1 \over r}
     \left( \tf{1}{2} + {1 \over 2 h} - {2 \over f} \right) - 
     {1 \over h r^2} \right] R = 0 \ .
  }
 A matching solution can be obtained in the usual fashion:
  \eqn\FMSoln{\eqalign{
   R_{\bf I} &= {h^{-i \lambda} \over 1-h} 
     F(-1,-\tf{1}{2} - 2 i \lambda,\tf{1}{2} - 2 i \lambda; h)  \cr
    &= {h^{-i \lambda} \over 1-h} \left( 1 + 
     {1 + 4 i \lambda \over 1 - 4 i \lambda} h \right)  \cr
   R_{\bf II} &= C_1 {1+h \over 1-h} + C_2 {\sqrt{h} \over 1-h}  \cr
   R_{\bf III} &= 2a \, e^{-i \rho} \left( 1 - {i \over 2\rho} \right) +
     ib {e^{i \rho} \over \rho}
  }}
 with 
  \eqn\MatchF{\eqalign{
   &C_1 = {1 \over 1 - 4 i \lambda} \qquad C_2 = 0  \cr
   &a = b = {i \over \omega r_0} {1 \over 1 - 4 i \lambda}
  }}
 where as before $\lambda = \omega / (4 \pi T)$.  The absorption
probability and cross-section are
  \eqn\FPandSigma{
   P = {(\omega r_0)^2 \over 4} (1 + 16 \lambda^2) \qquad\quad 
   \sigma_{\rm abs} = {2 \pi \over \omega^2} P = 
    {\pi r_0^2 \over 2} (1 + 16 \lambda^2) \ .
  }
 The $\lambda \to 0$ limit of \FPandSigma\ agrees with the general
result of \cite{dgm}.

\newsec{The effective string model}
\seclab\EffStr

The usual approach to effective string calculations (see for example
\cite{dmOne,cgkt,gkOne,krt}) has been to derive couplings between bulk
fields and the effective string by expanding the Dirac-Born-Infeld
(DBI) action to some appropriate order.  The leading terms in the
expansion specify a conformal field theory (CFT) which would describe
the effective string in the absence of interactions with bulk fields.
The interactions are dictated at leading order by terms in the
expansion which are linear in the bulk fields: for a scalar field
$\phi$, a typical coupling would be
  \eqn\CoupTyp{
   S_{\rm int} = \int d^2 x \, \phi(t,x,\vec{x}\!=\!0) \O(t,x)
  }
 where $\O(t,x)$ is some local conformal operator in the CFT.  The
integration is over the effective string world-volume, and $\vec{x}$
is set to zero because this is the location of the effective string in
transverse space.  One can then consider tree level processes mediated
by $S_{\rm int}$ where a bulk particle is converted into excitations
on the effective string.  Although the applicability of the DBI action
to D-brane bound states can be called into question, the prescription
described here for computing absorption or emission rates appears to
be very robust.  In the spirit of \cite{ja}, where effective string
calculations were used to account for properties of Reissner-Nordstrom
black holes without reference to any underlying microscopic picture,
let us examine the consequences of couplings of the general form
\CoupTyp.

Consider the absorption of a quanta of $\phi$ with energy $\omega$,
momentum $p$ along the effective string, and transverse momentum
$\vec{p}$.  I shall continue to use ${+}{-}{-}{-}{-}$ signature, so
for example $p \cdot x = \omega t - p x - \vec{p} \cdot \vec{x}$.  The
absorption cross-section can be calculated by setting
$\phi(t,x,\vec{x}\!=\!0) = e^{-i p \cdot x}$ and then treating
\CoupTyp\ as a time-dependent perturbation to the CFT which describes
the effective string in isolation.  Stimulated emission would be
calculated by choosing $e^{i p \cdot x}$ rather than $e^{-i p \cdot
x}$.  The $t$ and $x$ dependence of $\O(t,x)$ is fixed by the free
theory:
  \eqn\FreeEv{
   \O(t,x) = e^{i \hat{p} \cdot x} \O(0,0) 
    e^{-i \hat{p} \cdot x}
  }
 where $\hat{p} \cdot x = Ht - Px$, $H$ and $P$ being the Hamiltonian
and momentum operators of the CFT.  If one considers the perturbation
\CoupTyp\ to act for a time $t$, then Fermi's Golden Rule gives the
thermally averaged transition probability as
  \eqn\FermiGR{
   {\cal P} = \sum_{i,f} {e^{-\beta \cdot p_i} \over Z} P_{i \to f} = 
    Lt \sum_{i,f} {e^{-\beta \cdot p_i} \over Z} 
     (2 \pi)^2 \delta^2(p + p_i - p_f) 
      \left| \langle f | \O(0,0) | i \rangle \right|^2 \ .
  }
 This formula is valid for when the length $L$ of the effective string
is much larger than the Compton wavelength of the incoming scalar.
In \FermiGR, $\beta$ has two components, $\beta^+ = \beta_L$ and
$\beta^- = \beta_R$.  The partition function splits into left and
right sectors:
  \eqn\PartDef{
   Z = \tr e^{-\beta \cdot \hat{p}} = 
    (\tr_L e^{-\beta_L \hat{p}_+}) (\tr_R e^{-\beta_R \hat{p}_-}) \ .
  }
 For simplicity I take the momentum $p = (\omega,0,\vec{p})$ of the
incoming particle perpendicular to the brane, but clearly \FermiGR\
remains valid for the case of particles with Kaluza-Klein charge.

The summation over final states can become tedious when the scalar
turns into more than two excitations on the effective string.  Already
in the case of fixed scalars \cite{cgkt}, which split into two
right-movers and two left-movers, the evaluation of this summation was
a nontrivial exercise.  It therefore seems worthwhile to develop
further a method employed in \cite{ja} in which the absorption
probability is read off from the two point function of the operator
$\O$ in the effective string CFT.

Allow $t$ to take on complex values, defining $\O(t,x)$ by \FreeEv\
for arbitrary complex $t$.  According to usual notational conventions
\cite{Fetter}, $\O\+(t,x)$ is no longer the adjoint of $\O(t,x)$
except when $t$ is real; instead, $\O\+(t,x)$ is evolved from
$\O\+(0,0)$ using \FreeEv.  The conventional thermal Green's function
takes $t = -i \tau$ where $\tau$ is the Euclidean time:
  \eqn\TGreen{
   {\cal G}(-i \tau,x) = \langle \O\+(-i \tau,x) \O(0,0) \rangle = 
    \tr \left( \rho \,
     T_\tau \left\{ \O\+(-i \tau,x) \O(0,0) \right\} \right)
  }
 where $\rho = e^{-\beta \cdot \hat{p}} / Z$.  One can continue to
arbitrary complex $t$, defining $T_\tau$ to time-order with respect to
$-\Im(t)$.  The convenience of doing this is that the integral
  \eqn\GreenGotcha{
   \int d^2 x \, e^{i p \cdot x} {\cal G}(t - i \epsilon,x) = 
    \sum_{i,f} {e^{-\beta \cdot p_i} \over Z} 
     (2 \pi)^2 \delta^2(p + p_i - p_f) 
      \left| \langle f | \O(0,0) | i \rangle \right|^2
  }
 reproduces the right-hand side of \FermiGR.  The proof of
\GreenGotcha\ proceeds by inserting $\sum_i |i\rangle \langle i|$ and
$\sum_f |f\rangle \langle f|$ into \TGreen\ before $\O\+$ and $\O$
respectively.

Now let us turn to the evaluation of ${\cal G}(t,x)$.  Assume that
$\O(t,x)$ has the form
  \eqn\OForm{
   \O(t,x) = \O_+(x^+) \O_-(x^-)
  }
 where $x^\pm = t \pm x$ and $\O_+$ and $\O_-$ are primary fields of
dimensions $h_L$ and $h_R$, respectively.  Set $z = i x^-$ so that, for
$x$ real and $t = -i \tau$ imaginary, $\bar{z} = i x^+$.  The
singularities in ${\cal G}(t,x)$ are determined by the OPE's of $\O_+$
and $\O_-$ with themselves:
  \eqn\RLOPEs{\eqalign{
   \O_+(\bar{z}) \O_+\+(\bar{w}) &= 
    {C_{\O_+} \over (\bar{z}-\bar{w})^{2 h_L}} + 
     \hbox{less singular}  \cr
   \O_-(z) \O_-\+(w) &= 
    {C_{\O_-} \over (z-w)^{2 h_R}} + \hbox{less singular.}
  }} 
 ${\cal G}(t,x)$ factors into a left-moving and right-moving piece.
The imaginary time periodicity properties of each piece, together with
their singularities, suffice to fix the form of ${\cal G}(t,x)$
completely:\foot{Actually, there is a subtlety here: the information
from periodicity and singularities must be supplemented by a sum rule
\cite{Fetter} on the spectral density to squeeze out an ambiguity in
the analytic continuation.}
  \eqn\TPForm{
   {\cal G}(t,x) = {C_\O \over i^{2 h_L + 2 h_R}}
    \left( {\pi T_L \over \sinh \pi T_L x^+} \right)^{2 h_L}
    \left( {\pi T_R \over \sinh \pi T_R x^-} \right)^{2 h_R} 
  }
 where $C_\O = C_{\O_+} C_{\O_-}$.

At nonzero temperature, the absorption cross-section cannot be
calculated straight from \FermiGR: for bosons, the stimulated emission
probability must be subtracted off in order to obtain a result
consistent with detailed balance, as described in \cite{cgkt}.  The
net result is to set $\sigma_{\rm abs} {\cal F} t = {\cal P} (1 -
e^{-\beta \cdot p})$ where ${\cal F}$ is the flux and ${\cal P}$ is
read off from \FermiGR.  For fermions, the presence of an incoming
wave inhibits by the Exclusion Principle emission processes leading to
another fermion in the same state as the incoming wave.  The
absorption cross-section must therefore be calculated using
$\sigma_{\rm abs} {\cal F} t = {\cal P} (1 + e^{-\beta \cdot p})$.
Again, this result is in accord with detailed balance.

The considerations of the previous paragraph can be restated compactly
in terms of the Green's function: 
  \eqn\SigmaForm{\eqalign{
   \sigma_{\rm abs} &= {L \over {\cal F}} 
    \int d^2 x \, \big( {\cal G}(t-i\epsilon,x) - 
      {\cal G}(t+i\epsilon,x) \big)  \cr
    &= {L C_\O \over {\cal F}} 
       {(2 \pi T_L)^{2h_L-1} (2 \pi T_R)^{2h_R-1} \over 
        \Gamma(2h_L) \Gamma(2h_R)}
       {e^{\beta \cdot p / 2} - 
        (-1)^{2 h_L + 2 h_R} e^{-\beta \cdot p / 2} \over 2}  \cr
    &\qquad\cdot
       \left| \Gamma \left( h_L + i {p_+ \over 2 \pi T_L} \right)
              \Gamma \left( h_R + i {p_- \over 2 \pi T_R} \right)
       \right|^2 \ .
  }}
 One way to perform the integral in the first line is first to
separate into $x^+$ and $x^-$ factors and then to deform the contours
in the separate factors by setting $\epsilon = \beta_L/2$ or
$\beta_R/2$.  Assuming that bosons and fermions couple, respectively,
to conformal fields with $h_L + h_R$ an integer or half an odd
integer, one indeed obtains the factor $1 \mp e^{-\beta \cdot p}$
required by detailed balance.

The formula \SigmaForm\ represents almost the most general functional
form for an absorption cross-section that the effective string model
is capable of predicting.  One possible generalization is for the bulk
field to couple to a sum of different operators $\O(t,x)$, in which
case a sum of terms like \SigmaForm\ would be expected.  Another
generalization can arise from a coupling of a bulk field $\phi$ to the
effective string not through its value $\phi(t,x,\vec{x}\!=\!0)$ on
the string, as shown in \CoupTyp, but rather through its derivatives:
for instance $\partial_i \phi(t,x,\vec{x}\!=\!0)$ where $i$ labels a
transverse dimension.  In case of fields without Kaluza-Klein charge,
the effect of $n$ such derivatives is simply to introduce an extra
factor $\omega^{2n}$ on the right hand side of \SigmaForm.  Since the
flux is ${\cal F} = \omega$ for a canonically normalized scalar, the
$\omega$ dependence of the cross-section is 
  \eqn\SigmaZE{ 
   \sigma_{\rm abs} \sim \omega^{2n-1} 
    \sinh \left( \omega \over 2 T_H \right) 
   \left| \Gamma \left( h_L + i {\omega \over 4 \pi T_L} \right)
          \Gamma \left( h_R + i {\omega \over 4 \pi T_R} \right) 
    \right|^2 \ .
  }
 As we shall see in a specific example below, the flux factor for
massless fermions cancels out a similar factor in $C_\O$.  So for a
fermionic field which couples to the effective string through a term
in the lagrangian of the form $\partial^n \psi \O$, the energy
dependence of the cross-section is
  \eqn\FermionGrey{
   \sigma_{\rm abs} \sim \omega^{2n} 
      \cosh \left( \omega \over 2 T_H \right) 
    \left| \Gamma\left( h_L + i {\omega \over 4 \pi T_L} \right)
           \Gamma\left( h_R + i {\omega \over 4 \pi T_R} \right)
    \right|^2 \ .
  }

The remarkable fact is that numerous classical absorption calculations
that have appeared in the literature
\cite{dmw,dmOne,dmTwo,gkOne,gkTwo,cgkt,dkt,htr,kvk,km,krt,mast,ja,%
clOne,clTwo,dgm} all give results consistent with \SigmaForm\ or
\SigmaZE\ in the near-extremal limit.  As an example, consider
massless minimally coupled scalars falling into the four-dimensional
black hole considered in \cite{gkTwo}, whose effective string model is
derived from the picture of three intersecting sets of M5-branes.  The
absorption cross-section for the $\ell^{\rm th}$ partial wave is
  \eqn\ScalarEll{
   \sigma_{\rm abs}^\ell = {2 \over \omega^2}
    {(\omega T_H A_{\rm h})^{2 \ell + 1} \over (2 \ell)!^2
     (2 \ell + 1)!!} \sinh \left( \omega \over 2 T_H \right) 
     \left| \Gamma \left( \ell + 1 + i {\omega \over 4 \pi T_L} \right)
            \Gamma \left( \ell + 1 + i {\omega \over 4 \pi T_R} \right)
     \right|^2 \ ,
  }
 consistent with a coupling to the effective string of the form
$(\partial^\ell \phi) \O$ where $\O$ has dimensions $h_L = h_R = \ell
+ 1$.  Another interesting example is the fixed scalar
\cite{cgkt,kkOne}.  The cross-section for the $s$-wave in the
four-dimensional case \cite{kkOne}, with three charges equal and much
greater than the fourth ($R = r_1 = r_2 = r_3 \gg r_K \sim r_0$) is
  \eqn\FixedS{
   \sigma_{\rm abs} = {r_0^3 \over 2 \omega R^2} 
    \sinh \left( \omega \over 2 T_H \right) 
     \left| \Gamma \left( 2 + i {\omega \over 4 \pi T_L} \right)
            \Gamma \left( 2 + i {\omega \over 4 \pi T_R} \right)
     \right|^2 \ ,
  }
 consistent with a coupling of the form $\phi T_{++} T_{--}$.  It is
not hard to convince oneself that the analysis of the two-point
function works the same when $\O(t,x) = T_{++}(x^+) T_{--}(x^-)$ as it
did when $\O(t,x)$ was the product of left and right moving primary
fields.

One naturally expects that an equal charge black hole whose metric is
extreme Reissner-Nordstrom, like the $N=4$ example I focused on in
sections \MinCoup, \Prob.\Examples, and \OtherParticles, can be
obtained in the effective string picture by taking $T_L \gg
T_R,\omega$.  Indeed, it was found in \cite{ja} that ordinary scalar
cross-sections in this metric, and even in the Kerr-Newman metric,
have precisely the form one would expect from an effective string with
$T_L \gg T_R,\omega$.  The cross-sections have no dependence on $T_L$
in this limit, and the authors of \cite{ja} suggested a model which
made no reference to the left-moving sector.  Note however that
left-movers seem the natural explanation for the finite entropy of
extremal black holes---a subject not addressed in \cite{ja}.  In
\cite{hlm} it was argued in the context of $N=8$ compactifications
that the CFT on the effective string is a $(0,4)$ theory with central
charges $c_L = c_R = 6$.  The (local) $SU(2)$ $R$-symmetry of the
right-moving sector was identified with the group $SO(3)$ of spatial
rotations of the black hole.  The same identification of a local
$SU(2)$ on the effective string with $SO(3)$ was used in \cite{ja}.  I
will assume that an effective string description with $4$
supersymmetries and $c=6$ in the right-moving sector also applies to
the equal charge black hole of $N=4$ supergravity.

Without committing to specific assumptions about the nature or
existence of left-movers, one can conclude that the general form for
an effective string absorption cross-section of massless particles is
  \eqn\OneSided{\vcenter{\openup1\jot
    \halign{\strut\span\TT \quad& \span\TL & \span\TR & \quad\span\TT\cr
    bosons: & \sigma_{\rm abs} &\sim \omega^{2n} 
      \left| \Gamma(h_R + 2 i \lambda) \over 
             \Gamma(1 + 2 i \lambda) \right|^2 = 
      \omega^{2n} \prod_{r=1}^{h_R-1} (r^2 + 4 \lambda^2)
     & if $h_R \in {\bf Z}$  \cr
    fermions: & \sigma_{\rm abs} &\sim \omega^{2n} 
      \left| \Gamma(h_R + 2 i \lambda) \over 
             \Gamma({1\over 2} + 2 i \lambda) \right|^2 = 
      \omega^{2n} \prod_{r=1/2}^{h_R-1/2} (r^2 + 4 \lambda^2)
     & if $h_R \in {\bf Z}+\tf{1}{2}$  \cr
  }}}
 where $r$ runs over integers in the case of bosons and integers plus
$1/2$ in the case of fermions, and 
  \eqn\LambdaAgain{
   \lambda = {\omega \over 4 \pi T} = {\omega \over 8 \pi T_R} \ .
  }
 The cross-section \NonExProb\ for minimally coupled photons fits the
form \OneSided\ with $h_R=2$ and $n=1$.  The form of the coupling of
the minimal photon to the effective string is further constrained by
rotation invariance.  If we represent the right-moving sector using
$4$ free chiral bosons (which are neutral under $SU(2)$) and an
$SU(2)$ doublet of fermions, then simplest coupling to the minimal
photon with the right group theoretic properties is
  \eqn\PhotonCoupling{
   {\cal L}_{\rm int} = 
    \phi_{\alpha\beta} \Psi_-^\alpha \partial_- \Psi_-^\beta F_+ + \hc 
  }
 where $F_+$ is a left-moving field which, as noted previously, does
not affect the form of the absorption cross-section.  Recall that
$\phi_{\alpha\beta}$ is a field strength: one derivative is hidden
inside it, so indeed $n=1$.  The indices on $\Psi_-^\alpha$ are for
the group $SO(3)$ of spatial rotations, while the indices on
$\phi_{\alpha\beta}$ are for the $SU(2)_L$ half of the Lorentz group
$SO(3,1)$.  But they can be contracted as shown in a static gauge
description since the generators of $SO(3)$ are just sums of the
generators of $SU(2)_L$ and $SU(2)_R$.  With the current spinor index
conventions, an upper dotted index is equivalent to a lower undotted
index if only spatial $SO(3)$ rotations are considered.

The dilaton and axion cross-sections \AFProbs\ fit the form \OneSided\
with $h_R = 2$, and so does the minimal fermion cross-section with
$h_R = 3/2$.  The natural guess is a coupling of these fields to the
stress-energy tensor and the supercurrents: to linear order in all the
fields,
  \eqn\SuperCoupling{
   {\cal L}_{\rm int} = \big[ (\phi + i B) T_{--} +
     \Lambda_{3\alpha} T_{F-}^\alpha -
     i \Lambda_{4\alpha} T_{F-}^{\dagger\alpha} \big] F_+ + \hc
  }
 The form of \SuperCoupling\ is dictated by the quarter of the $N=4$
supersymmetry which is preserved by the extreme black hole.  The terms
in the supersymmetry variation of ${\cal L}_{\rm int}$ with no
derivatives can shown to cancel using
  \eqn\SuperVars{\eqalign{
   \delta (\phi + i B) &= \epsilon^\alpha \Lambda_{3\alpha} +
    \bar\epsilon_{\dot\alpha} i \sqrt{2} \sigma^{0\dot\alpha\beta}
     \Lambda_{4\beta}  \cr
   \delta T_{F-}^\alpha &= \epsilon^\alpha T_{--} \ .
  }}
 Here $\epsilon_\alpha$ parameterizes what was referred to in
\cite{klopp} as the $\epsilon^{34}_+$ supersymmetry.  Note that in the
conventions outlined in the appendix, both $\sqrt{2}
\sigma^{0\dot\alpha\beta}$ and $\sqrt{2} \sigma^0_{\alpha\dot\beta}$
are numerically the identity matrix.  The left-moving sector of the
CFT is assumed to be neutral under supersymmetry.

The equivalence (in static gauge) of upper undotted and lower dotted
indices has been used to simplify \SuperCoupling.  The matrices
$\sqrt{2} \sigma^{0\dot\alpha\beta}$ and $\sqrt{2}
\sigma^0_{\alpha\dot\beta}$ can be used to convert between them.  The
index on $T_{F-}\+$ has been raised in \SuperCoupling\ using
$\epsilon^{\alpha\beta}$.  A consequence of the second line in
\SuperVars\ is thus $\delta T_{F-}^{\dagger\alpha} =
-\bar\epsilon_{\dot\alpha} T_{--}$.

 The normalization of $T_{F-}^\alpha$ used here differs from \cite{ss}
by a factor of $\sqrt{2}$: if $G_n^\alpha$ and $L_n$ are the
supercurrent and Virasoro generators of \cite{ss}, then the present
conventions are to set $T_{F-}^\alpha(z) = {1 \over \sqrt{2}} \sum_n
z^{-n-3/2} G_n^\alpha$ and $T(z) = \sum_n z^{-n-2} L_n$.  On the
complex plane, the nonzero two point functions are 
  \eqn\HolTwoPt{\eqalign{
   \langle T_{F-}^\alpha(z) T_{F-\beta}\+(w) \rangle &= 
    {c \over 3} {\delta^\alpha_\beta \over (z-w)^3}  \cr
   \langle T_{--}(z) T_{--}(w) \rangle &= {c/2 \over (z-w)^4} \ .
  }}
 I will also assume that the only nonzero two point function of $F_+$
and $F_+\+$ is
  \eqn\RightTwoPt{
   \langle F(\bar{z}) F\+(\bar{w}) \rangle = 
    {C_F \over (\bar{z} - \bar{w})^{2 h_R}} \ .
  }

Now we are ready to compute effective string cross-sections.  For the
dilaton, the operator $\O(t,x)$ entering into the analysis of
\CoupTyp-\SigmaForm\ is $\O(t,x) = T_{--}(x^-) \big(F_+(x^+) +
F_+\+(x^+) \big)$.  The fact that $T_{--}$ is not primary does not
alter the periodicity properties of its two-point function.  Thus the
arguments leading from \OForm\ to \SigmaForm\ still apply, and $C_\O =
c C_F = 6 C_F$.  Because of the non-canonical normalization of the
dilaton field in the action \FullBL, the particle flux in a wave $\phi
= e^{-i p \cdot x}$ is ${\cal F} = 4 \omega$, four times the usual
value.  Plugging these numbers into \SigmaForm\ and taking the large
$T_L$ limit, one obtains
  \eqn\DilCross{
   \sigma_{\rm abs} = {L C_F \over 8 T} (2 \pi T_L)^{2 h_L - 1}
    (2 \pi T_R)^3 {\Gamma(h_L)^2 \over \Gamma(2 h_L)}
    (1 + 4 \lambda^2) \ .
  }
 The axion of course yields the same result.

The minimal fermions clearly have the same cross-section, so let us
consider only $\Lambda_3$.  Let the incoming wave be
$\Lambda_{3\alpha} = u_\alpha e^{-i p \cdot x}$.  With $\O(t,x) =
u_\alpha T_{F-}^\alpha(x^-) F_+(x^+)$, the analysis leading to
\SigmaForm\ goes through as usual, yielding
  \eqn\CEval{
   C_\O = \bar{u}_{\dot\alpha} \delta_\alpha^\beta u_\beta 
      {c C_F \over 3}  
    = \bar{u}_{\dot\alpha} \sqrt{2} \sigma^{0\dot\alpha\beta} u_\beta \,
       2 C_F \ .
  }
 In the second equality the equivalence of lower undotted and upper
dotted indices has again been used.  Recall that $\sqrt{2}
\sigma^{0\dot\alpha\beta}$ is indeed the identity matrix.  The flux is
${\cal F} = 4 \bar{u}_{\dot\alpha} \sqrt{2} \sigma^{0\dot\alpha\beta}
u_\beta$.  (As for the dilaton, the $4$ here is due to the
non-canonical normalization of the fermion field: see the footnote at
the beginning of section~\OtherParticles.\Fermions.)  Now \SigmaForm\
can be used again to give
  \eqn\FermCross{
   \sigma_{\rm abs} = {\pi L C_F \over 16} (2 \pi T_L)^{2 h_L - 1}
    (2 \pi T_R)^2 {\Gamma(h_L)^2 \over \Gamma(2 h_L)}
    (1 + 16 \lambda^2) \ .
  }
 The effective string cross-sections \DilCross\ and \FermCross\ stand
in the same ratio as the semi-classical cross-sections for fixed
scalars and minimal fermions quoted in \AFProbs.

Of course, it would be highly desirable to carry out calculations
similar to the ones presented here for the black holes in five
dimensions that can be modelled using the D1-brane D5-brane bound
state.  There one can hope that an understanding of the soliton
picture can fix overall normalizations; but again one expects the
residual supersymmetry to fix relative normalizations between
fermionic and bosonic cross-sections.

\newsec{Conclusion}
\seclab\Conclusion

One of the main technical results of this paper has been to show that
when the equations of motion for photons or fermions are simple enough
to be analyzed by the dyadic index methods of
\cite{teuk,teukI,Churil}, they lead to ordinary differential equations
whose near-horizon form is hypergeometric.  That fact alone gives
their low-energy absorption cross-sections a form which is capable of
explanation in the effective string description.

Dyadic index methods are not essential to the analysis of minimally
coupled fermions; indeed, the Weyl equation for fermions has been
analyzed recently in \cite{dgm} in arbitrary dimensions using more
conventional techniques.  However, the dyadic index method provides an
efficient, unified treatment of minimal fermions and minimal photons.
Indeed, the photon radial equations \ExactExtremeI\ and \ExNonEx\ seem
difficult to derive by other means.  The existence of minimally
coupled photons seems to depend essentially on the equal charge
condition, which makes the dilaton background constant.  Minimally
coupled fermions, as I suggested in section~\OtherParticles.\Fermions,
may be more common because their existence depends on the vanishing of
their supersymmetry variations in the black hole background.

The greybody factors computed in \NonExProb\ and \AFProbs\ are
polynomials in the energy $\omega$ rather than quotients of gamma
functions as found in \cite{mast}.  This is characteristic of an
effective string whose left-moving temperature $T_L$ is much greater
than $T_R$ and $\omega$.  These polynomial greybody factors are
sufficient to determine the conformal dimension of the right-moving
factor in the operator through which a field couples to the effective
string, and the number of derivatives in that coupling.  For example,
the minimal photon couples through its field strength times a $h_R=2$
operator.  But \NonExProb\ and \AFProbs\ do not yield any information
regarding the left-movers.  To see the effects of left-movers, one
might try to generalize the present treatment to black holes far from
extremality, as was done recently in \cite{clTwo} for minimally
coupled scalars.

The absence of a string soliton description of the equal charge black
hole in pure $d=4$, $N=4$ supergravity precludes a precise comparison
of cross-sections between the effective string and semi-classical
descriptions.  However, by assuming that the effective string
world-sheet theory is a $(0,4)$ super-conformal field theory whose
right-moving $R$-symmetry group, $SU(2)$, is identified with the group
of spatial rotations, it has been possible to show that the relative
normalizations of the dilaton, axion, and minimally coupled fermion
cross-sections are correctly predicted by the effective string.  The
proposed couplings of these fields to the effective string are
simple: the scalars couple to the stress-energy tensor while the
fermions couple to the supercurrent.  One would expect that it
possible to extend this picture to a manifestly supersymmetric
specification of how all the massless bulk fields couple to the
effective string at linear order.

There is a simple point which nevertheless is worth emphasizing: the
cross-sections of the dilaton, axion, and minimal fermions are related
by supersymmetry despite their different energy dependence.  The
energy dependence (also known as the greybody factor) arises from
finite-temperature kinematics.  Unsurprisingly, the kinematic factors
are different for particles of different spin; but their form turns
out to be fixed by the conformal dimension of the field by which a
field couples to the effective string.  Supersymmetry acts on the
$S$-matrix, relating the coefficients I have called $C_\O$ in
section~\EffStr.  The predictions of supersymmetry regarding the
absorption cross-sections of different particles in the same multiplet
thus have more to do with the relative normalization than the energy
dependence.

\bigbreak\bigskip\bigskip\centerline{{\bf Acknowledgements}}\nobreak

I would like to thank C.~Callan, S.~Das, G.~Horowitz, S.~Mathur,
A.~Peet, A.~Strominger, and particularly I.R.~Klebanov for useful
discussions.  This work was supported in part by DOE grant
DE-FG02-91ER40671, the NSF Presidential Young Investigator Award
PHY-9157482, and the James S.~McDonnell Foundation grant 91-48.  I
also thank the Hertz Foundation for its support.

\vfill\eject


\appendix{A}{Dyadic index conventions}

This appendix presents in a pedestrian fashion the aspects of the
Newman-Penrose formalism relevant to the rest of the paper.  A
readable introduction can be found in \cite{Wald}; for an authoritative
treatment the reader is referred to \cite{pr}.

Sign conventions vary by author, and the ones used here are as close
as possible to those of the original paper by Newman and Penrose
\cite{np} and to those of Teukolsky \cite{teuk,teukI}.  First 
consider flat Minkowski spacetime with
mostly minus metric, $\eta_{ab} =
\diag(1,-1,-1,-1)$.  The conventions on raising and lowering spinor
indices are those of ``northwest contraction:''
  \eqn\NW{\vcenter{\openup1\jot
    \halign{\strut\span\TL & \span\TR & \span\TT & \span\TL & \span\TR\cr
   \psi^\alpha &= \epsilon^{\alpha\beta} \psi_\beta &\qquad&
    \psi_\alpha &= \psi^\beta \epsilon_{\beta\alpha} \cr
   \bar\psi^{\dot\alpha} &= 
     \epsilon^{\dot\alpha\dot\beta} \bar\psi_{\dot\beta} &\qquad&
    \bar\psi_{\dot\alpha} &= 
     \bar\psi^{\dot\beta} \epsilon_{\dot\beta\dot\alpha} \cr
  }}}
 where the sign of the antisymmetric tensors is fixed by
  $\epsilon_{01} = \epsilon^{01} = 
   \epsilon_{\dot{0}\dot{1}} = \epsilon^{\dot{0}\dot{1}} = 1$.
 In flat space, the conventional choice of the matrices 
$\sigma^a_{\alpha\dot\beta}$ which map bispinors to vectors is 
  \eqn\sigmas{
   \sigma^a = {1 \over \sqrt{2}} 
    \left( 1,\tau_3,\tau_1,-\tau_2 \right) 
  }
 where the matrices $\tau_i$ are the standard Pauli matrices.  
Vector indices are interchanged with pairs of spinor indices using 
the formulae
  \eqn\VectorSpinor{
   v^a = \sigma^a_{\alpha\dot\alpha} v^{\alpha\dot\alpha} \qquad
   v^{\alpha\dot\alpha} = \sigma_a^{\alpha\dot\alpha} v_a 
  } 
 where 
  $\sigma_a^{\alpha\dot\alpha} = \eta_{ab} \epsilon^{\alpha\beta}
   \epsilon^{\dot\alpha\dot\beta} \sigma^b_{\beta\dot\beta}$,
 consistent with our northwest contraction rules.  
 There is no need to define matrices $\bar\sigma^{a\dot\alpha\beta}$. 
The metric has a simple form when written with spinor indices:
  \eqn\MetricForm{
   \eta_{ab} \sigma^a_{\alpha\dot\alpha} \sigma^b_{\beta\dot\beta} =
    \epsilon_{\alpha\beta} \epsilon_{\dot\alpha\dot\beta} \ .
  }

In curved spacetime, the metric $g_{\mu\nu}$ is again chosen with
${+}{-}{-}{-}$ signature.  In this paper, the metric is always of the
form
  \eqn\DiagonalMetric{
   ds^2 = e^{2A(r)} dt^2 - e^{2B(r)} dr^2 - 
    e^{2C(r)} \left( d\theta^2 + \sin^2 \theta d\phi^2 \right) \ .
  }
 It will turn out to be useful to define not only the standard
diagonal vierbein
  $e_\mu^a = \diag(\sqrt{g_{tt}},\sqrt{-g_{rr}},\sqrt{-g_{\theta\theta}},
   \sqrt{-g_{\phi\phi}})$,
 but also a complex null tetrad\foot{In the literature it is common to
see factors of $g_{tt}$ included in the definitions of $\ell^\mu$ and
$n^\mu$ so that seven rather than six of the spin coefficients vanish.
This however complicates the time-reversal properties of the
solutions.}
  \eqn\ComplexNullTetrad{\vcenter{\openup1\jot
    \halign{\strut\span\TL & \span\TR & \span\TT & \span\TL &
     \span\TR\cr
   \ell^\mu &= {e^\mu_t + e^\mu_r \over \sqrt{2}} &\qquad&
    n^\mu &= {e^\mu_t - e^\mu_r \over \sqrt{2}} \cr
   m^\mu &= {e^\mu_\theta + i e^\mu_\phi \over \sqrt{2}} &\qquad&
    \bar{m}^\mu &= {e^\mu_\theta - i e^\mu_\phi \over \sqrt{2}} \ . \cr
  }}}
 One of the conveniences of working with spinors is that a spinor is a 
sort of square root of a null vector: for any spinor $\psi^\alpha$, 
  $v^\mu = e^\mu_a \sigma^a_{\alpha\dot\alpha} 
   \psi^\alpha \bar\psi^{\dot\alpha}$ 
is a null vector, and any null vector can be written in this form.  It is
important to note that in the context of the Newman-Penrose formalism,
spinor components are ordinary commuting numbers, not Grassmann numbers.
It is possible to introduce a basis $(\omicron^\alpha,\iota^\alpha)$ 
for spinor space with the properties
  \eqn\SpinorBasis{\vcenter{\openup1\jot
    \halign{\strut\span\TC\cr
   \omicron_\alpha \iota^\alpha = 1 \cr
   \ell^{\alpha\dot\alpha} = \omicron^\alpha \bar\omicron^{\dot\alpha}
    \quad
   n^{\alpha\dot\alpha} = \iota^\alpha \bar\iota^{\dot\alpha} \quad
   m^{\alpha\dot\alpha} = \omicron^\alpha \bar\iota^{\dot\alpha} \quad
   \bar{m}^{\alpha\dot\alpha} = \iota^\alpha \bar\omicron^{\dot\alpha}
    \ . \cr
  }}}
 A particular choice of $(\omicron^\alpha,\iota^\alpha)$ is
  \eqn\OmicronIota{
   \omicron^\alpha = \pmatrix{1 \cr 0 } \qquad
   \iota^\alpha = \pmatrix{0 \cr 1 } \ .
  }
 Dyadic indices are introduced by defining 
  $\xi_0^\alpha = \omicron^\alpha$,
  $\xi_1^\alpha = \iota^\alpha$   
 and writing $\psi_\Gamma$ for the components of the spinor
$\psi_\alpha$ with respect to the basis $\xi_\Gamma^\alpha$:
  \eqn\DyadSum{
   \psi_\Gamma = \xi_\Gamma^\alpha \psi_\alpha \qquad
   \psi_\alpha = -\xi^\Gamma_\alpha \psi_\Gamma \ .
  }
 The minus sign in the second equation is the result of insisting on 
the same raising and lowering conventions for dyadic indices as for 
spinor indices: 
  $\xi^\Gamma_\alpha = \epsilon^{\Gamma\Delta} \xi_\Delta^\beta 
   \epsilon_{\beta\alpha}$.

It is a familiar story \cite{bd} how the minimal $SO(3,1)$
connection $\omega_\mu{}^a{}_b$ on the local Lorentz bundle is induced 
from the Christoffel connection: one defines
  \eqn\SpinConnection{
   \omega_\mu{}^a{}_b = e^a_\nu \partial_\mu e^b_\nu + 
    e^a_\nu \Gamma^\nu_{\mu\rho} e_b^\rho 
  }
 so that 
  \eqn\Consist{
   \nabla_\mu v^a = \partial_\mu v^a + \omega_\mu{}^a{}_b v^b
     = e^a_\nu \nabla_\mu v^\nu 
     = e^a_\nu (\partial_\mu v^\nu + \Gamma^\nu_{\mu\rho} v^\rho) \ .
  }
 The Newman-Penrose spin coefficients are defined in an exactly
analogous way.  In fact, they are merely special (complex) linear
combinations of the $\omega_\mu{}^a{}_b$.  It is conventional in the
literature to make dyadic indices ``neutral'' under the covariant
derivative $\nabla_\mu$: $\nabla_\mu v_\Gamma = \partial_\mu v_\Gamma$.  The
covariant derivatives of spinors, by contrast, are defined using the
connection induced from $\omega_\mu{}^a{}_b$.  It is convenient to
define a ``completely covariant'' derivative $D_\mu$ and a connection
$\gamma_\mu{}^\Sigma{}_\Gamma$ with the defining properties
  \eqn\DerDef{
   D_\mu \psi_\Gamma = 
    \partial_\mu \psi_\Gamma - \psi_\Sigma \gamma_\mu{}^\Sigma{}_\Gamma = 
    \xi_\Gamma^\alpha \nabla_\mu \psi_\alpha \ . 
  }
 A brief way of characterizing the covariant derivative is to say that
under $\nabla_\mu$, the quantities $g_{\mu\nu}$, $\eta_{ab}$,
$\epsilon_{\alpha\beta}$, $e^a_\mu$, and $\sigma^a_{\alpha\dot\alpha}$
(together with their alternative incarnations $\eta^{ab}$,
$\epsilon^{\dot\alpha\dot\beta}$, etc.) are covariantly constant.  Under
$D_\mu$, the quantities $\xi_\Gamma^\alpha$ are covariantly constant as
well.  From \DerDef\ it is immediate that\foot{Note that I choose
the sign for $\gamma_{\Delta\dot\Delta\Gamma\Sigma}$ according to the
convention of \cite{teuk} and \cite{np} rather than of \cite{Wald}.}
  \eqn\SpinCoefDef{
   \gamma_{m\Sigma\Gamma} = 
    -\xi_{\Gamma\alpha} \nabla_\mu \xi_\Sigma^\alpha \ . 
  }
 The quantities 
  $\gamma_{\Delta\dot\Delta\Gamma\Sigma} = 
   \sigma^\mu_{\Delta\dot\Delta} \gamma_{m\Gamma\Sigma}$
 are the spin coefficients.  They have the symmetry 
  $\gamma_{\Delta\dot\Delta\Gamma\Sigma} = 
   \gamma_{\Delta\dot\Delta\Sigma\Gamma}$.
 With twelve independent complex components they represent the same
information as the forty real components of the Christoffel connection
$\Gamma^\mu_{\nu\rho}$.  It is useful to note that
  \eqn\SigmaDyad{
   \sigma^\mu_{\Delta\dot\Delta} = 
    e^\mu_a \xi_\Delta^\alpha \bar{\xi}_{\dot\Delta}^{\dot\alpha}
     \sigma^a_{\alpha\dot\alpha} = 
    \pmatrix{ \ell^\mu & m^\mu \cr
              \bar{m}^\mu & n^\mu } \ .
  }
 A useful formula for calculating the spin coefficients can be given 
in terms of $\sigma^\mu_{\Delta\dot\Delta}$:
  \eqn\EasyDef{
   \gamma_{\Delta\dot\Delta\Gamma\Sigma} = 
    -\tf{1}{2} \xi_\Gamma^\beta \bar\xi_{\dot\Gamma}^{\dot\beta} 
     \sigma^\nu_{\Delta\dot\Delta} \nabla_\nu 
     ( \xi_{\Sigma\beta} \bar\xi^{\dot\Gamma}_{\dot\beta} ) = 
    -\tf{1}{2} \sigma^\mu_{\Gamma\dot\Gamma}
     \sigma^\nu_{\Delta\dot\Delta} \nabla_\nu
     \sigma_{\mu\Sigma}{}^{\dot\Gamma} \ .
  }
 Some further notational definitions are conventional in dyadic index
papers:
  \eqn\DyadNotation{\vcenter{\openup1\jot
    \halign{\strut\span\TL & \span\TR & \span\TT & \span\TL & \span\TR\cr
   \gamma_{0\dot{0}\Gamma\Sigma} &= 
    \pmatrix{ \kappa & \epsilon \cr \epsilon & \pi } &\qquad&
   \gamma_{0\dot{1}\Gamma\Sigma} &=
    \pmatrix{ \sigma & \beta \cr \beta & \mu } \cr
   \gamma_{1\dot{0}\Gamma\Sigma} &=
    \pmatrix{ \rho & \alpha \cr \alpha & \lambda } &\qquad&
   \gamma_{1\dot{1}\Gamma\Sigma} &=
    \pmatrix{ \tau & \gamma \cr \gamma & \nu } \cr
  }}}
  \eqn\DerNotation{
   D = \ell^\mu \nabla_\mu \qquad \Delta = n^\mu \nabla_\mu \qquad
   \delta = m^\mu \nabla_\mu \qquad \bar\delta = \bar{m}^\mu \nabla_\mu \ .
  }
 For the metric \DiagonalMetric, one finds
  \eqn\DiagonalSC{\eqalign{
    \kappa &= \pi = \sigma = \lambda = \tau = \nu = 0 \cr
    \epsilon &= \gamma = {e^{-B} A' \over 2 \sqrt{2}} \qquad
      \beta = -\alpha = {e^{-C} \cot \theta \over 2 \sqrt{2}} \qquad
      \mu = \rho = -{e^{-B} C' \over \sqrt{2}} 
  }}
 where primes denote derivatives with respect to $r$.  \DiagonalSC\
represents a remarkably economical way of describing the connection of
an arbitrary spherically symmetric spacetime: there are only three
independent nonzero spin coefficients, and they are real.

\vfill\eject
\centerline{\bf References}
\bibliography{/usr/people/ssgubser/papers/bib/brane}   
\bibliographystyle{/usr/people/ssgubser/tex/bibtex/ssg}   

\bye